\documentclass[a4paper,11pt]{article}
\usepackage{jheppub} % for details on the use of the package, please

\usepackage{graphicx}
\usepackage{amsmath}
\usepackage{amssymb}
\usepackage{comment}
\usepackage{hyperref}

\usepackage{subfigure}

\newcommand*{\rn}{\ensuremath{{\mathbf{r}_0}}}
\renewcommand*{\l}{\ensuremath{{\mathbf{l}}}}
\renewcommand*{\v}{\ensuremath{{\mathbf{v}}}}
\newcommand*{\p}{\ensuremath{{\mathbf{p}}}}
\renewcommand*{\d}{\ensuremath{{\mathrm{d}}}}
\renewcommand*{\i}{\ensuremath{{\mathrm{i}}}}
\renewcommand*{\mod}{\ensuremath{ \ {\mathrm{mod} \ }}}

\makeatletter
\newcommand*\wt[1]{\mathpalette\wthelper{#1}}
\newcommand*\wthelper[2]{%
    \hbox{\dimen@\accentfontxheight#1%
        \accentfontxheight#11.15\dimen@
        $\m@th#1\widetilde{#2}$%
        \accentfontxheight#1\dimen@
    }%
}
\newcommand*\accentfontxheight[1]{%
    \fontdimen5\ifx#1\displaystyle
        \textfont
    \else\ifx#1\textstyle
        \textfont
    \else\ifx#1\scriptstyle
        \scriptfont
    \else
        \scriptscriptfont
    \fi\fi\fi3
}
\makeatother

\author[a]{Zhaoyu Han}
\affiliation[a]{Department of Physics, Stanford University, Stanford, CA 94305, USA}
\author[b]{and Jing-Yuan Chen}
\affiliation[b]{Institute for Advanced Study, Tsinghua University, Beijing, 100084, China}

\title{Fractional Hall Conductivity and Spin-c Structure in \\ Solvable Lattice Hamiltonians}
\date{\today}

\abstract{
The Kapustin-Fidkowski no-go theorem forbids $U(1)$ symmetric topological orders with non-trivial Hall conductivity in (2+1)d from admitting commuting projector Hamiltonians, where the latter is the paradigmatic method to construct exactly solvable lattice models for topological orders. Even if a topological order would intrinsically have admitted commuting projector Hamiltonians, the theorem forbids so once its interplay with $U(1)$ global symmetry which generates Hall conductivity is taken into consideration. Nonetheless, in this work, we show that for all (2+1)d $U(1)$ symmetric abelian topological orders of such kind, we can construct a lattice Hamiltonian that is controllably solvable at low energies, even though not ``exactly'' solvable; hence, this no-go theorem does not lead to significant difficulty in the lattice study of these topological orders. Moreover, for the fermionic topological orders in our construction, we introduce the lattice notion of spin-c structure -- a concept important in the continuum that has previously not been adequately introduced in the lattice context.}

\begin{document}

\maketitle
\flushbottom

\section{Introduction}
\label{sect_intro}

Since the surprising discovery of the quantum Hall effect, topological phases of matter have become an important subject in condensed matter physics. An approach of central importance in the theoretical study of gapped topological phases of matter is the construction of exactly solvable lattice models, in particular the models of commuting projector Hamiltonians \cite{Kitaev:1997wr,Levin:2004mi}. There are obvious favorable gains from this approach -- for instance, an explicit understanding of the exotic topological properties, and a proof of principle of the microscopic realizability (which is also seen as a possible route towards quantum computation \cite{Kitaev:1997wr}). But there is a higher layer of reasons why this approach is important: The construction of such exactly solvable models is not at all a gymnastics of smart tricks; rather, there are deep principles that allow such models to be systematically constructed. Understanding these principles is, in itself, a pinnacle in our understanding of the physics of gapped topological phases and the mathematics behind them.

Generally speaking, in the now well-established paradigm, the construction of commuting projector lattice Hamiltonians for a gapped topological phase relies on a gapped boundary condition of that topological phase \cite{Kitaev:2011dxc}. Not all gapped topological phases support a gapped boundary, but when one does, starting with the topological properties of its boundary excitations, we have systematic machinery to generate a commuting projector lattice Hamiltonian \cite{Kirillov:2011mk, Bhardwaj:2016clt}, and the physical picture behind this seemingly involved machinery is, as natural as one may think of, that of coarse-graining renormalization, till reaching the fixed-point \cite{Levin:2004mi, Bhardwaj:2016clt}. Roughly speaking, the role of having a gapped boundary condition is so that we have good control over what happens at the boundary between two neighboring ``grains'' at low energies, then we view the ``grains'' as effective lattice sites, and the fixed-point self-consistency conditions as ``rules of the game'', in order to construct the lattice model.

An obvious question then arises. There are many interesting gapped topological phases that only support gapless boundary conditions but not gapped ones. They are therefore beyond the current standard approach of constructing commuting projector Hamiltonians. What do we understand about those topological phases? For example, can we find -- not by hand-wavy arguments, but on a firm basis -- lattice models that realize those topological phases?  This is the general theme of problems that our current paper aims to explore. The goal of our paper is to address the problem of lattice realization for a certain class of systems, in hope that some novel features that arise in our work may shed light on the more general developments of methods that will extend the current paradigm.

For example, systems in (2+1)-dimensions with non-trivial Hall conductivity -- the very property that led to the first experimental discovery of topological phases -- are, somewhat ironically, not within the reach of the established paradigm of commuting projector Hamiltonian. Indeed, these systems only support gapless boundary conditions that transport electric charge in a chiral manner, in order to conserve electric charge in the presence of a Hall conductivity in the bulk \cite{halperin1982quantized, Wen:1995qn}. Here, we can see the involvement of a global symmetry, the $U(1)$ for charge conservation. More broadly, the notion of a \emph{symmetry enriched} topological order involves both an intrinsic topological order (which concerns the braiding and fusion of anyons) as well as its interplay with a global symmetry, making it more interesting than just the intrinsic topological order itself; the standard approach of commuting projector lattice Hamiltonian has also been extended to incorporate global symmetries \cite{Barkeshli:2014cna, heinrich2016symmetry, cheng2017exactly}. When symmetry enrichment is taken into consideration, even if an intrinsic topological order may originally admit gapped boundary conditions, under certain ways of enrichments, those gapped boundary conditions may violate the global symmetry on the boundary and hence become prohibited. The prohibition of gapped boundary conditions then prohibits commuting projector Hamiltonian realizations. This scenario indeed happens in systems with Hall conductivity. It has been rigorously proven by Kapustin and Fidkowski \cite{kapustin2020local} (see also \cite{PhysRevB.105.L081103}) that, under mild physical assumptions, Hall conductivity is impossible in any gapped commuting projector Hamiltonian model. The no-go theorem applies regardless of whether the intrinsic topological order admits a commuting projector lattice Hamiltonian, as long as the $U(1)$ global symmetry enrichment is in place.

We focus on the context of $U(1)$ global symmetry and Hall conductivity for the following reasons. First, the $U(1)$ global symmetry of electric charge conservation is perhaps the most fundamental symmetry in condensed matter physics, and the coupling to the associated electromagnetic field (as a background field) and in particular the Hall conductivity response provide physically important probes to a system of interest. Second, $U(1)$ global symmetry is a continuous symmetry, as opposed to the discrete symmetries that are more often studied in the exactly solvable lattice model context. Third, the primary examples of intrinsic topological orders coupled to $U(1)$ global symmetry are abelian topological orders, which are relatively simple and hence suitable for exploring new ideas beyond the established framework.

Our main result is that, for any bosonic or fermionic abelian topological order in (2+1)d that intrinsically admits a commuting projector Hamiltonian realization, even if such realization becomes prohibited after a $U(1)$ enrichment that generates Hall conductivity, we can still construct a lattice Hamiltonian that can be controllably solved at low energies, and show the Hamiltonian realizes the desired topological order with the desired Hall conductivity. (In our previous work \cite{PhysRevB.105.155130}, we studied the simplest among the bosonic cases.) Moreover, for the fermionic cases, our construction leads to a simple and natural notion of \emph{spin-c} structure on the lattice. The spin-c structure, which roughly speaking ``mixes'' the $U(1)$ background field and the spin structure (fermion boundary conditions), is an important notion for studying fermionic systems in the continuum \cite{Seiberg:2016rsg}, but has yet to be adequately introduced in the lattice context. 

We would like to remark that our construction is vastly different from a model for free fermion Chern insulators, but closer to exactly solvable models such as the toric code. Firstly, in terms of the results, we can realize fractional Hall conductivities while free fermion models only realize integer ones. Yet a more fundamental difference is that all physics in our construction occurs at the lattice scale, while the Hall physics in a free fermion model is only found in the infrared. More exactly, in our construction we can solve the theory under arbitrary $U(1)$ background on the lattice, while in a free fermion model the Hall response can only be computed for perturbatively small $U(1)$ backgrounds at long wavelengths.

In the next section, we will give a more comprehensive theoretic overview of the background and the ideas of our work. For now, we would like to emphasize the main new feature that arises in our construction, compared to the current paradigmatic framework which does not apply to our cases of interest. In our construction, a low energy ``string type'' (in the language of \cite{Levin:2004mi}) is no longer tied to a particular state in the local full Hilbert space. Rather, the association of a ``string type'' to a local state only emerges dynamically at low energies (in a controlled manner), in a way that may depend on the local background data. Such a more flexible relation between a ``string type'' and a local state may be generically useful in future theoretical developments on the (symmetry enriched) topological orders beyond the current paradigm.

\section{Theoretic overview}
\label{sect_intro2}

A remarkable fact about abelian topological orders in (2+1)d is that they, as well as their add-on aspects such as symmetry enrichment and spin-c compatibility, can be all captured by abelian Chern-Simons theories in the continuum \cite{Belov:2005ze}. This will provide important guidance on how to construct our lattice theories and solve them.

We are particularly interested in those abelian topological orders which admit commuting projector constructions in the absence of $U(1)$ global symmetry enrichment, but cease to be so, due to the Kapustin-Fidkowski no-go theorem~\cite{kapustin2020local, PhysRevB.105.L081103}, after $U(1)$ enrichment in certain ways that generate a non-trivial Hall conductivity. In the continuum Chern-Simons description, such theories \emph{essentially} (see below) take the ``doubled'' form such that there are two dynamical gauge fields, with one appearing as a Lagrange multiplier:
\begin{align}
S = \int d^3 x \: \left[ \frac{n}{2\pi} a d b + \frac{k}{4\pi} bdb - \frac{q}{2\pi} Ad a - \frac{p}{2\pi} A d b \right]
\label{doubleCS}
\end{align}
with $n, k, p, q$ integers. Here $a, b$ are dynamical $U(1)$ gauge fields and $A$ is the ``electromagnetic'' $U(1)$ background for the $U(1)$ global symmetry. (We use upper case letters for background fields and lower case letters for dynamical fields.) When the coupling to $A$ is absent, integrating out the Lagrange multiplier-like gauge field $a$ will reduce the theory of the $b$ field to a $\mathbb{Z}_n$ gauge theory with Dijkgraaf-Witten twist $k$~\cite{Kapustin:2010hk, Seiberg:2016rsg}, therefore $n$ and $k$ determine the intrinsic topological order, with $n=1$ topologically trivial~\cite{Witten:2003ya}. (For generic values of $k$, the theory is only defined in oriented spacetime. In the below we will assume the orientability of the spacetime.) On the other hand, the electric charges $p, q$ determine the $U(1)$ global symmetry enrichment. (In principle, $a, b$ can couple to two different $U(1)$ global symmetry backgrounds $A, B$, but in this work, we will identify $B=A$.) When the integers $k, p, q$ are subjected to further conditions, more interesting aspects of the theory -- such as spin and spin-c -- can be discussed, as we will see soon. This theory has a Hall conductivity of $-2pq/n+q^2 k/n^2$; it supports two elementary types of anyons, one with unit coupling to $a$ and the other to $b$, that carry electric charges $-kq/n^2 + p/n$ and $q/n$ respectively; they also have $\pi k /n^2$ and $0$ self-statistics respectively, and $-2\pi/n$ mutual statistics~\cite{Seiberg:2016rsg, Wen:1995qn}. These are abelian anyons, so any integer linear combination of them is again an anyon, with the electric charge and statistical phases linear combined accordingly.

By the word ``essentially'' above \eqref{doubleCS}, we mean that the most general cases of interest are such that the integers $n$ and $k$ become integer square matrices \cite{Kapustin:2010hk}, and $q$ and $p$ become integer vectors. But such generalization of \eqref{doubleCS} does not lead to substantial changes in the main points of this paper, so we shall just work with the theories described by \eqref{doubleCS}.

There is an important distinction between even and odd values of $k$~\cite{Seiberg:2016rsg, Wen:1995qn}. The theories with even $k$ are already well-defined with the apparent bosonic degrees of freedom. On the other hand, the odd $k$ theories are implicitly fermionic -- it turns out an intrinsic fermion must be attached to each $2\pi$ flux excitation of $b$ (which is a certain linear combination of the aforementioned anyons) in order for the theory to be well-defined. This requires extra background data on the spacetime since we thereby have to specify the spin structure -- i.e. the fermion boundary conditions being anti-periodic or periodic.

The situation becomes more interesting when both the $U(1)$ global symmetry enrichment and the bosonic versus fermionic issue are simultaneously considered. A theory is said to satisfy the \emph{spin-charge condition} if the bosons always have even electric charges while the fermions always have odd charges \cite{Wen:1995qn, Seiberg:2016rsg}. If this spin-charge condition is indeed satisfied, then the theory may be defined with fewer background data -- it can be defined in a \emph{spin-c} background~\cite{Seiberg:2016rsg} instead of a $U(1)$ background together with a spin structure. In three spacetime dimensions or lower, where any oriented manifold admits spin structure(s), a spin-c background is particularly simple to describe: When the effects of a change of the spin structure (the fermion boundary conditions) can always be compensated by introducing a flat $\pi$ holonomy of the $U(1)$ background -- which is true if and only if the spin-charge condition is satisfied -- we may view this equivalence relation as a redundancy in the background data, and we thus define a spin-c background as the $U(1)$ background plus the spin structure modulo this equivalence relation. Condensed matter systems generally satisfy the spin-charge condition, because, in the end, the only relevant microscopic dynamical degree of freedom is electron, which is a fermion with an odd electric charge, therefore the microscopic theory does admit the said equivalence relation, and any low energy effective description must also do. (Including neutrons, for instance, will violate the spin-charge condition; however neutrons are usually not included in the dynamics of condensed matter systems.) For systems of our interest, whose effective theories are of the form \eqref{doubleCS}, satisfying the spin-charge condition means $q=0 \mod 2$ and $p=k\mod 2$, since the $2\pi$ flux excitation of $a$ (which is an anyon with coupling $n$ to $b$) is always bosonic and is of electric charge $q$, while the $2\pi$ flux excitation of $b$ (which is an anyon with couplings $n$ to $a$ and $k$ to $b$) is bosonic/fermionic when $k$ is even/odd and is of electric charge $p$ \cite{Wen:1995qn}.

Our goal is to realize the $U(1)$ symmetric abelian topological theories in \eqref{doubleCS} by solvable Hamiltonians on the lattice. When the $U(1)$ global symmetry enrichment is absent, such theories reduce to $\mathbb{Z}_n$ gauge theories with Dijkgraaf-Witten twist $k$, which have well-known commuting projector Hamiltonian realizations in both the bosonic~\cite{Levin:2004mi, mesaros2013classification, hu2013twisted, Lin:2014aca} and fermionic~\cite{Gu:2013gma, Gu:2012ib, Gaiotto:2015zta, Tarantino:2016qfy, Bhardwaj:2016clt} cases; we reproduce these familiar results in Appendix \ref{app:descrete_model}. As we take the $U(1)$ global symmetry into consideration, some of the enrichments will retain the admission of commuting projector Hamiltonians; this is when the $p, q$ values are such that the Hall conductivity vanishes. In this regard, see \cite{Levin:2011hq} for an early example, \cite{Wang:2021smv} for a complete discussion of the bosonic cases, and \cite{wang2021exactly} for some fermionic cases. A more interesting scenario is when the $U(1)$ enrichment is such that the Hall conductivity is non-zero. Commuting projector Hamiltonian realization is no longer possible, according to the Kapustin-Fidkowski no-go theorem \cite{kapustin2020local,PhysRevB.105.L081103}. For these cases, the best hope is to construct lattice Hamiltonians that can be solved in a controlled manner at low energies, such that the universal properties, most particularly the Hall conductivity, can be explicitly demonstrated. 

In our previous work~\cite{PhysRevB.105.155130}, using the intuition provided by the continuum double $U(1)$ Chern-Simons theory \eqref{doubleCS}, the untwisted theories with $k=0$ have been constructed and solved, showing that controllably solvable lattice Hamiltonians can indeed be constructed for theories with Hall conductivity. (The lattice Hamiltonian has also been suggested in \cite{geraedts2013exact} based on different reasoning, although the parameters considered there are different from the solvable limit in \cite{PhysRevB.105.155130}.) Here we show the same can be completed for general values of $k$, including the particularly interesting odd values that are associated with fermionic theories. This means:
\begin{quote}
\emph{For any (2+1)d abelian topological order which intrinsically admits a commuting projector lattice Hamiltonian (admits a gapped boundary condition), after it is enriched by $U(1)$ global symmetry, even if a non-trivial Hall conductivity is generated (so that the $U(1)$ symmetry prohibits any gapped boundary condition), we can always construct a lattice Hamiltonian controllably solvable at low energies.}  
\end{quote}
Moreover, we may choose the parameters of our Hamiltonian so that it is, in a loose sense as we shall see, ``arbitrarily close to being exactly solvable''.

Our construction is based on the use of a Villainized version of the \emph{dynamical} $U(1)$ gauge fields, as has been explored in \cite{Chen:2019mjw} in details for the present purpose. In the Villainized $U(1)$ gauge theory, by introducing on the lattice plaquettes an integer variable subjected to a $1$-form $\mathbb{Z}$ gauge invariance, the $2\pi$ flux excitations of the $U(1)$ gauge field are explicitly and locally accounted for, hence effectively capturing all the topological aspects of a $U(1)$ gauge theory, such as non-trivial bundles and large gauge transformations, through the central extension sequence $2\pi\mathbb{Z}\rightarrow\mathbb{R}\rightarrow U(1)$. In a more general language that has become increasingly important nowadays~\cite{pfeiffer2003higher, Gukov:2013zka}, we shall say the gauge group $U(1)$ is elevated to a gauge 2-group $[U(1)/\mathbb{R}]$. On the other hand, the \emph{background} $U(1)$ field, i.e. the electromagnetic field $A$, must be presented as an ordinary $U(1)$ on the lattice -- we cannot ``choose'' it to take the Villainized form in our construction because it is a background data; to put in other words, the $U(1)$ global symmetry is not a $[U(1)/\mathbb{R}]$ 2-group global symmetry. This difference might seem subtle, but it will, in the end, lead to the introduction of non-commuting terms in the lattice Hamiltonian~\cite{Chen:2019mjw, PhysRevB.105.155130}, thereby circumventing the Kapustin-Fidkowski no-go theorem. As we proceed, we will see how this comes about in detail.

More broadly speaking, in this paper, to accommodate for the most general $U(1)$ global symmetry enrichments in the abelian topological orders of interest, we need to embed the (twisted) $\mathbb{Z}_n$ gauge field in the intrinsic topological order into the central extension sequence $2\pi\mathbb{Z} \rightarrow \mathbb{R} \rightarrow U(1)$ (up to a $n/2\pi$ rescaling between the $\mathbb{Z}_n$ and the $U(1)$) with a shifted origin. Here and below the first term in the sequence is a $1$-form symmetry to be gauged. To accommodate for only those $U(1)$ global symmetry enrichments without Hall conductivity, the origin no longer needs to be shifted, and the central extension sequence can be reduced to $n\mathbb{Z} \rightarrow \mathbb{Z} \rightarrow \mathbb{Z}_n$, which is what is used in \cite{Wang:2021smv}, see Appendix~\ref{app:descrete_model}. When the $U(1)$ global symmetry is ignored, a general twisted $\mathbb{Z}_n$ intrinsic topological order uses the central extension sequence $n\mathbb{Z}_n \rightarrow \mathbb{Z}_{n^2} \rightarrow \mathbb{Z}_n$, which is what is used in \cite{PRXQuantum.3.010353}, see Appendix~\ref{app:descrete_model} also; meanwhile an untwisted theory, i.e. the original $\mathbb{Z}_n$ toric code \cite{Kitaev:1997wr}, can use the trivial central extension sequence $1 \rightarrow \mathbb{Z}_{n} \rightarrow \mathbb{Z}_n$, which is just $\mathbb{Z}_n$ outright.

Finally, through our construction, for the odd $k$ fermionic theories we automatically obtain a lattice notion of spin-c background. As we introduced above, in three dimensions or lower, where any oriented manifold admits spin structure(s), the continuum notion of spin-c background is a $U(1)$ background data plus a spin structure data modulo a global equivalence relation. On the lattice, the spin structure, which is defined globally in the continuum, is encoded into a local $\mathbb{Z}_2$ data~\cite{Gu:2012ib, Gu:2013gma, Gaiotto:2015zta, Tarantino:2016qfy, Bhardwaj:2016clt}. We find that the global equivalence relation that appears in the spin-c definition also becomes local on the lattice. Hence the lattice notion of spin-c background in (2+1)d can be phrased in a completely local manner: 
\begin{quote}
    \emph{In 2d spatial lattice Hamiltonian theories, the spin-c background data is such that the local $\mathbb{Z}_2$ spin structure data on each individual link can be absorbed as a $\pi$ shift into the $U(1)$ background data.}
\end{quote}
This definition applies to 3d spatial lattice as well since any 3d oriented manifold admits spin structure(s) too. (In higher dimensions, it is easy to envision that the lattice spin-c definition must still be stated locally instead of globally, but this is beyond the scope of the present paper.) As far as we know, such a notion, though simple and natural as it is, has previously not been adequately introduced in the lattice context, therefore it is useful for us to state it explicitly here. (A Villainized version of lattice spin-c has been introduced in \cite{Chen:2019mjw} in (2+1)d and in \cite{Kobayashi:2021jsc} in (3+1)d. However, such Villainized version of lattice spin-c can only be defined in global terms, as opposed to the local terms here, and, similar to Villainized global $U(1)$ we discussed before, is only suitable for effective spacetime lattice Lagrangians but not for actual spatial lattice Hamiltonians. See Section 3 of \cite{Chen:2019mjw} for discussions.)

This completes our overview of the theoretical background, motivation and main ideas of our work. In the following sections, we will first introduce the bosonic and fermionic Hamiltonians and the ideas behind their constructions, and then sketch how we solve them controllably (with details in the appendices), and finally compute their topological properties, in particular the Hall conductivity.

\section{Hamiltonian}
\label{sect_Hamiltonian}

In this section we introduce our lattice Hamiltonian construction, motivated by the continuum gauge theory \eqref{doubleCS}. The necessary intermediate ideas, which involve an effective lattice Lagrangian, were explored in \cite{Chen:2019mjw} and summarized in Appendix \ref{app_L_to_H}. (If we disregard the $U(1)$ global symmetry, the effective lattice Lagrangian can be readily reduced to the familiar exactly solvable $\mathbb{Z}_n$ Dijkgraaf-Witten Hamiltonians \cite{Levin:2004mi, mesaros2013classification, hu2013twisted, Lin:2014aca}, see Appendix \ref{app:descrete_model}; moreover, the discussion there is closely related to \cite{PRXQuantum.3.010353}.) 

For now, we assume the spatial lattice is a square lattice for simplicity; in Sec.~\ref{sect_latticegeneralization} we will generalize all of our results to an arbitrary triangulated space with a branching structure.

\subsection{Bosonic theories}
\label{sect_boson}

\begin{figure}[t]
\centering
\raisebox{-0.5\height}{\subfigure[]{\label{fig_lattice}\includegraphics[width=.45\textwidth]{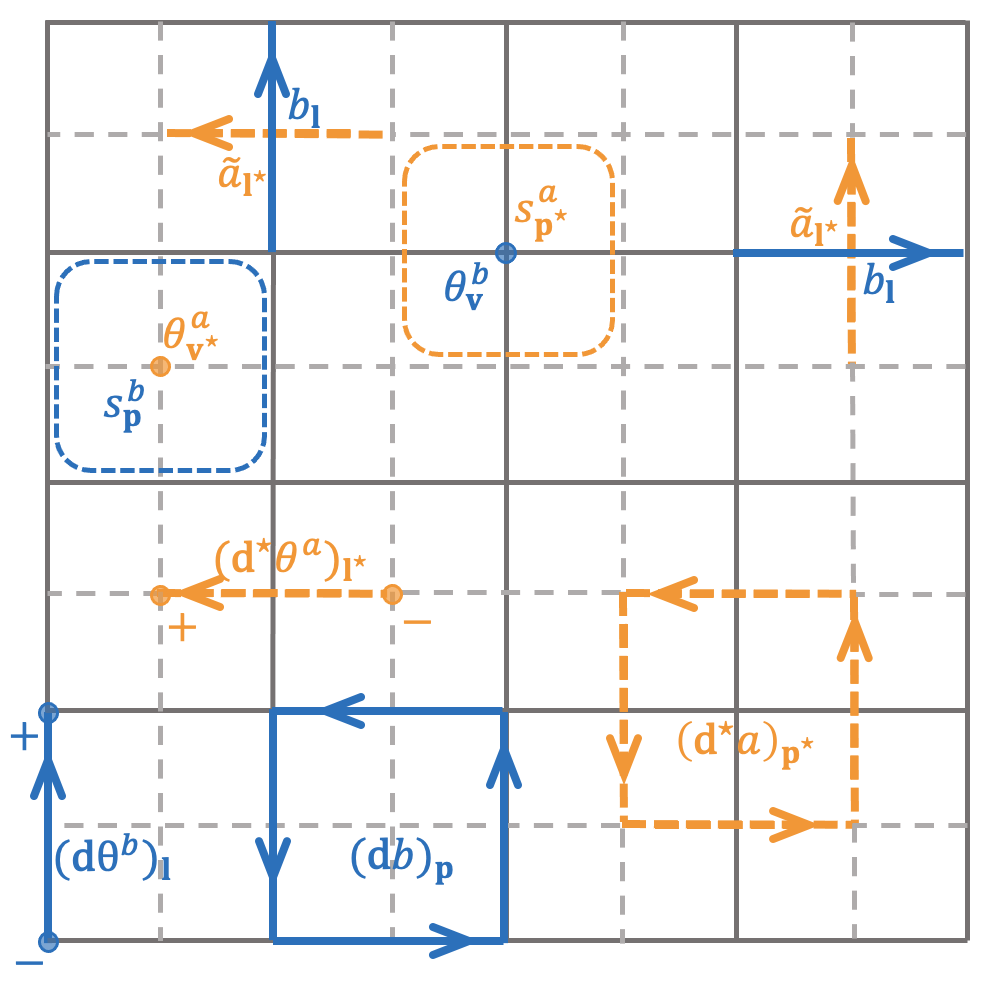}}}
\raisebox{-0.5\height}{\subfigure[]{\label{fig_lattice_A}\includegraphics[width=.45\textwidth]{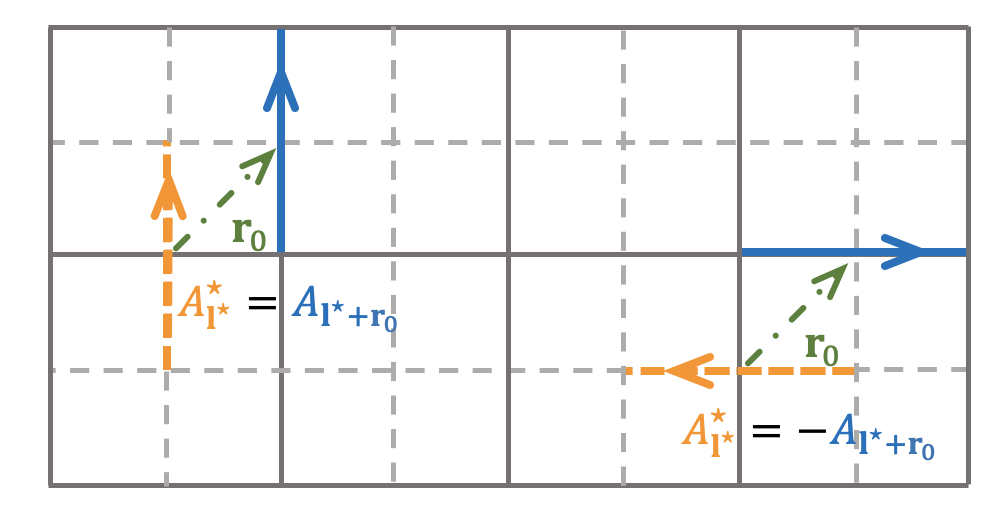}}}
\caption{(a) Illustration for the lattice (solid line) and the dual lattice (dash line), the conjugate pairs of operators, and the lattice and dual lattice ``exterior derivatives''. Note we have picked the natural directions of the lattice links to be $+\hat{\mathbf{x}}$, $+\hat{\mathbf{y}}$, and the natural directions of the dual lattice links to be $90$-degrees counter-clockwise to the associated lattice link directions, as labeled by the arrows. When the direction label on a link (or a dual link) reverses, the associated operator picks a negative sign. (b) A choice of identification of $A^\star_{\l^\star}$ with $A_\l$ through relation $A^\star_{\l^\star} = \eta_{\l^\star} A_{\l^\star+\rn}$, where $\l$ is chosen to locate at $\rn \equiv \hat{\mathbf{x}}/2+\hat{\mathbf{y}}/2$ away from $\l^\star$ and $\eta_{\l^\star}=\mp 1$ for dual link $\l^\star$ oriented in $-\hat{\mathbf{x}}$ or $\hat{\mathbf{y}}$ direction.}

\end{figure}

We first consider the bosonic theories with even $k$. The local Hilbert spaces and their endowed operators are (see Fig.~\ref{fig_lattice} for illustration):
\begin{itemize}
\item
On each link $\l$, which is equivalent to a dual lattice link $\l^\star$ (with direction 90 degrees counter-clockwise to the direction of $\l$), there is a local Hilbert space endowed with a conjugate pair of real-valued operators satisfying
\begin{align}
[b_{\l=\l^\star}, \tilde{a}_{\l^\star}] = \i \frac{2\pi}{n},
\label{ba_comm}
\end{align}
where $\tilde{a}_{\l^\star} \equiv a_{\l^\star} + \frac{k}{2n} \eta_{\l^\star} b_{\l^\star-\rn}$ is a local combination of the ``original'' lattice gauge fields $a$ and $b$, $\rn \equiv \hat{\mathbf{x}}/2+\hat{\mathbf{y}}/2$, and $\eta_{\l^\star}=\mp 1$ for dual link $\l^\star$ oriented in $-\hat{\mathbf{x}}$ or $\hat{\mathbf{y}}$ direction. (Note that $a_{\l^\star}$ and $a_{\l^\star+\rn}$ do not commute.)
\item
On each vertex $\v$, which is equivalent to a dual lattice plaquette $\p^\star$, there is a local Hilbert space endowed with a conjugate pair of integer/$U(1)$ valued operators satisfying
\begin{align}
[s^a_{\p^\star=\v}, e^{\i\theta^b_\v}] = e^{\i\theta^b_\v}.
\label{sathetab_comm}
\end{align}
\item
On each plaquette $\p$, which is equivalent to a dual lattice vertex $\v^\star$, there is a local Hilbert space endowed with a conjugate pair of integer/$U(1)$ valued operators satisfying
\begin{align}
[s^b_{\p=\v^\star}, e^{\i\theta^a_{\v^\star}}] = e^{\i\theta^a_{\v^\star}}.
\label{sbthetaa_comm}
\end{align}
\end{itemize}
Since a dual link $\l^\star$ is right on top of a link $\l$, we use $\l^\star=\l$ to indicate that $\l$ and $\l^\star$ are centered at the same position; notations $\v=\p^\star$ and $\p=\v^\star$ can be understood similarly. 

To have an intuitive picture in mind (which will appear to be an emergent picture that is not exact, after we introduce our Hamiltonian later), one may think of $a, b$  as dynamical $\mathbb{R}$ gauge fields on the lattice, such that they are effectively reduced to dynamical $U(1)$ gauge fields upon the introduction of the associated dynamical Dirac string variables $s^a$ and $s^b$. The ``reduction to $U(1)$'' can be understood in the following ``Villain'' sense. (In the more general theoretic language of higher groups \cite{pfeiffer2003higher, Gukov:2013zka}, the Villain realization of $U(1)$ is an action 2-group $[U(1)/\mathbb{R}]$.) The effective $U(1)$ fluxes associated with the dynamical gauge fields $a, b$ are
\begin{align}
f^b_\p \equiv (\d b - 2\pi s^b)_\p, \ \ \ \ \ f^a_{\p^\star} \equiv (\d^\star a - 2\pi s^a)_{\p^\star}
\label{generator_f}
\end{align}
where the lattice curl $\d b$ and the dual lattice curl $\d^\star a$ are illustrated in Fig.~\ref{fig_lattice}. Thanks to the Dirac string variables, if we sum $f^b_\p$ or $f^a_{\p^\star}$ over all the plaquettes $\p$ or $\p^\star$ on a closed space (consider a square lattice plane with periodic boundary conditions, forming a torus), we can have arbitrary $2\pi\mathbb{Z}$ values, reproducing the Dirac quantization condition for $U(1)$ gauge fluxes.

A major feature to be noted is the following $1$-form $\mathbb{Z}$ gauge invariances (a general introduction to higher form symmetries can be found in \cite{Gaiotto:2014kfa})
\begin{align}
& b_\l \: \rightarrow \: b_\l + 2\pi z^b_\l, \ \ \ \ \ s^b_\p \: \rightarrow \: s^b_\p + (\d z^b)_\p, \nonumber \\[.1cm]
& a_{\l^\star} \: \rightarrow \: a_{\l^\star} + 2\pi z^a_{\l^\star}, \ \ \ \ \ s^a_{\p^\star} \: \rightarrow \: s^a_{\p^\star} + (\d^\star z^a)_{\p^\star}
\label{1-form_Z}
\end{align}
where $z^b_\l$, $z^a_{\l^\star}$ are arbitrary $\mathbb{Z}$ valued transformations on the links and dual links respectively. Such invariances manifest the fact that $a, b$ are effectively reduced from $\mathbb{R}$ to $\mathbb{R}/2\pi\mathbb{Z} = U(1)$ gauge fields. The corresponding generators for these transformations are
\begin{align}
g^a_{\l^\star} &\equiv e^{\i(\d^\star\theta^a-na)_{\l^\star} - \i \frac{\eta_{\l^\star} k}{2}(b_{\l^\star+\rn} + b_{\l^\star-\rn})}\nonumber\\ 
g^b_{\l} &\equiv e^{\i(\d\theta^b-nb)_{\l}},
\label{generator_g}
\end{align}
where the notions of $\d\theta^b$ and $\d^\star\theta^a$ are illustrated in Fig.~\ref{fig_lattice}. The transformation $z^b_\l$ in \eqref{1-form_Z} is generated by conjugating $b_\l$ with $g^a_{\l^\star=\l}$. (Note that the application of $g^a_{\l^\star}$ keeps all $a_{\l^\star}$ invariant, not $\tilde{a}_{\l^\star}$.) Likewise, the transformation $z^a_{\l^\star}$ is generated by conjugating $a_{\l^\star}$ with $g^b_{\l =\l^\star }$. The expressions of $g^a, g^b$ invite us to think of $\theta^a, \theta^b$ as $U(1)$ non-linear sigma model phases (``superconducting phases''), that describe bosons $s^b$, $s^a$  with charge $n$ under effectively $U(1)$ dynamical gauge fields $a, b$, respectively; the $s^b$ bosons are further charged $k$ under $b$. 

Indeed, the generators above -- and supposedly the desired theory itself -- also have the ordinary ($0$-form) gauge transformations
\begin{align}
& b_\l \: \rightarrow \: b_\l + (\d\varphi^b)_\l, \ \ \ \ \ a_{\l^\star} \: \rightarrow \: a_{\l^\star} + (\d^\star\varphi^a)_{\l^\star},\nonumber \\
& \theta^b_\v \: \rightarrow \: \theta^b_\v + n\varphi^b_\v,\nonumber\\
&\theta^a_{\v^\star} \: \rightarrow \: \theta^a_{\v^\star} + n\varphi^a_{\v^\star} + \frac{k}{2} \left(\varphi^b_{\v^\star-\rn}+\varphi^b_{\v^\star+\rn} \right), 
\label{0-form_U1}
\end{align}
where $\varphi^b_\v$, $\varphi^a_{\v^\star}$ are arbitrary $U(1)$ transformations on the vertices and dual vertices respectively; they are effectively $U(1)$ instead of $\mathbb{R}$ because any $2\pi\mathbb{Z}$ part of them can be completely absorbed into \eqref{1-form_Z}. The two transformations by $\varphi^a_{\v^\star}$, $\varphi^b_\v$ are, in turn, generated by the commutators with
\begin{align}
    G^a_{\p^\star=\v} \equiv& f^a_{\p^\star} + \frac{k}{2n} \left(f^b_{\p^\star+\rn}+f^b_{\p^\star-\rn}\right) \nonumber\\
    G^b_{\p=\v^\star} \equiv& f^b_{\p}.
\end{align}
To generate the $0$-form effectively $U(1)$ gauge transformations, they act on $b, \theta^b$ and $a, \theta^a$ by commutation.

It is easy to check that the generators $g^a, g^b, G^a, G^b$ for the gauge transformations \eqref{1-form_Z} and \eqref{0-form_U1} all commute with each other, so in particular their own expressions are invariant under \eqref{1-form_Z} and \eqref{0-form_U1}. If a state is a simultaneous eigenstate of all of $g^a, g^b, G^a, G^b$, we may view the state as respecting the respective Gauss's law constraints with the ``gauge charges'' given by the simultaneous eigenvalues. Such states are the gauge invariant states. With this understanding of the nature of the constraints, it is intuitive to motivate the following ``prototype'' Hamiltonian, following the usual idea of imposing emergent gauge constraints using energy costs \cite{Kitaev:1997wr, Levin:2004mi}:
\begin{align}
\wt{H} \ =& \ \frac{U_b}{2}\sum_\l \left| 1 - e^{iqA_\l} g^b_\l \right|^2 + \frac{U_a}{2}\sum_{\l^\star} \left| 1 - e^{ipA^\star_{\l^\star}} g^a_{\l^\star} \right|^2 \nonumber \\[.2cm]
& + \ \frac{V_b}{2}\sum_\p \left(G^b_\p\right)^2 + \frac{V_a}{2} \sum_{\p^\star} \left(G^a_{\p^\star}\right)^2
\end{align}
where $A_\l$ is the electromagnetic $U(1)$ background field living on the link $\l$, while $A^\star_{\l^\star}$ on the dual link $\l^\star$ is identified with a nearby $A_\l$ pointing in the same direction, see Fig.~\ref{fig_lattice_A} for a choice of identification. The coupling of the electromagnetic field $A$ into the system can be intuitively understood as the following. If we think of $s^a_{\p^\star}$ as a boson number operator on the vertex $\v=\p^\star$, then $e^{\pm i\theta^b_\v}$ is the creation/annihilation operator of the boson, and hence $g^b_\l$ involves the hopping of such a boson across the link $\l$; if the boson carries electric charge $q$, its hopping will indeed couple to $A$ through the factor $e^{iqA_\l}$. Likewise for the term with $A^\star_{\l^\star}$. Therefore, the local electric charge operator on a vertex $\v$ is given by
\begin{align}
\rho_\v \ \equiv \ q \, s^a_{\p^\star=\v} + p \, s^b_{\p=\v-\rn}
\label{charge_density}
\end{align}
where the second term is due to the said identification between $A^\star_{\l^\star}$ and $A_\l$.

However, similar to what we have discussed in detail in \cite{PhysRevB.105.155130}, there are serious problems with such ``prototype'' Hamiltonian $\wt{H}$:
\begin{enumerate}
\item Since $g^a, g^b, G^a, G^b$ commute with each other and their eigenvalues are all continuous, the Hamiltonian is not a gapped one that we want. In fact, the Hamiltonian is locally gapless everywhere, not just becoming gapless only after taking the limit of large system size. Local gaplessness is highly unphysical (fine-tuned), and no universal physics can be possible. 
\item With a little extra effort we will see that, the ground state expectation values of physical observables have ambiguous results, somewhat like asking for the expectation $\langle x \rangle$ for a Bloch wavefunction $\exp(\i k x)$. The ambiguity arises from the infinite ranges of the local values of $a, b$ involved in the superposition.
\end{enumerate}
These two issues actually represent the same problem -- since $\tilde{a}, b$ are canonical conjugates, the local gaplessness of $\tilde{a}$ is related to the local unboundedness of $b$, and vice versa.

We thus modify the Hamiltonian to make it gapped and its $a, b$ values softly bounded, by adding simple, non-commuting terms $\epsilon_a \tilde{a}_{\l^\star}^2/2$ and $\epsilon_b b_{\l}^2/2$ to the ``prototype'' Hamiltonian $\wt{H}$. The total Hamiltonian then reads
\begin{align}\label{BosonHamiltonian}
&H = \wt{H}+\sum_{\l^\star} \frac{\epsilon_a}{2} \tilde{a}_{\l^\star}^2 + \sum_\l \frac{\epsilon_b}{2} b_\l^2  \nonumber \\[.1cm]
& \ \ \ =\sum_{\l} H^{(1)}_{\l}+\sum_{\l^\star} H^{(2)}_{\l^\star} + \sum_{\p} H^{(3)}_{\p}+ \sum_{\p^\star} H^{(4)}_{\p^\star} 
\end{align}
which we further decompose into four parts for our solving procedure later (see section \ref{sect_bosonic} for explicit expressions).

% \begin{align}
% &H^{(1)}_{\l} =  \frac{\epsilon_a}{2} \tilde{a}_{\l^\star}^2 - U_b \cos(\d\theta^b - nb + qA)_\l \nonumber\\ 
% &H^{(2)}_{\l^\star} = \frac{\epsilon_b}{2} b_\l^2  \nonumber \\
% & \ \ \ \ \ \ \ \ \   - U_a \cos\left[(\d^\star\theta^a - n\tilde{a} + pA^\star)_{\l^\star} - \frac{k}{2}\eta_{\l^\star}b_{\l^\star+\rn}\right]  \nonumber\\[.2cm]
% &H^{(3)}_{\p} = V_b \left( \d b - 2\pi s^b \right)_\p^2 \nonumber \\
% &H^{(4)}_{\p^\star} =  V_a\left[\left( \d^\star \tilde{a} - 2\pi s^a \right)_{\p^\star} -\frac{\pi k}{n}  s^b_{\p^\star-\rn} \right. \nonumber \\
% & \ \ \ \ \ \ \ \ \ \ \ \ \ \ \ \ \ \ \ \ \ \ \ \ \ \ \ \left. + \frac{k}{2n} (\d b -2\pi s^b)_{\p^\star+\rn} \right]^2 \ .
% \end{align}
% Recall $\rn \equiv \hat{\mathbf{x}}/2+\hat{\mathbf{y}}/2$, and $\eta_{\l^\star}=\mp 1$ for dual link $\l^\star$ oriented in $-\hat{\mathbf{x}}$ or $\hat{\mathbf{y}}$ direction, see Fig.~\ref{fig_lattice}; $A^\star$ is identified with $A$ through relation $A^\star_{\l^\star} = \eta_{\l^\star} A_{\l^\star+\rn}$, see Fig.~\ref{fig_lattice_A}. [Throughout this paper, if an object specified with original lattice position and an object specified with dual lattice position appear together, their positions should be understood as identical, e.g. the link $\l$ and dual link $\l^\star$ in $H^{(1)}_\l$ should be understood as centered at the same position, $\l=\l^\star$.]

Since the $\epsilon_a, \epsilon_b$ terms do not commute with some of the remaining terms (nor with each other), a gap is opened up; the fluctuations of $a, b$ on each link are also softly bounded by these terms. As the $\epsilon_a, \epsilon_b$ terms violate the gauge invariances \eqref{1-form_Z}, \eqref{0-form_U1}, the gauge field picture of the $a, b$ variables is no longer exact, but emergent at best. Indeed, in the remaining sections of this paper, we will show such a gauge field picture emerges at the low energy sector on each individual link when $\epsilon_a, \epsilon_b$ are small compared to $U_a, U_b$. We will solve the low energy sectors of the full Hamiltonian, and show the system exhibits the desired Hall conductivity.

\subsection{Fermionic theories and lattice \emph{spin-c} structure}
\label{sect_fermion}

It is well-known that when $k$ is odd, the desired topological order must contain intrinsic fermions, otherwise, there would be an anomalous $1$-form $\mathbb{Z}_2$ symmetry being gauged. Upon introducing the intrinsic fermions in a suitable way, the gauge anomaly is canceled, and the system would, in turn, have a global fermion parity $\mathbb{Z}_2^f$.

Let us see how to accommodate these facts in our construction. When $k$ is odd, the $g^a$ operators -- which generate the $z^b$ $1$-form $\mathbb{Z}$ gauge transformation in \eqref{1-form_Z} -- at $\l^\star$ and $\l^\star+\rn$ no longer commute, but anti-commute instead. This means the $1$-form $\mathbb{Z}_2$ gauge transformation parameterized by $z^b \mod 2$ becomes anomalous. In terms of the Hamiltonian, the $U_a$ terms in the Hamiltonian \eqref{BosonHamiltonian} at links separated by $\rn$ anti-commute rather than commute, ruining the ``almost commuting'' nice property of $H^{(2)}$. (Following the idea of Appendix F of \cite{PhysRevB.105.155130}, it is not hard to argue that such a Hamiltonian has a topologically trivial low energy sector, which is indeed expected for a theory whose supposed ``emergent gauge field'' turns out anomalous and hence ceases to be a gauge field.)
Below we will explain how to suitably couple the odd $k$ Hamiltonian to fermions in order to obtain the desired fermionic topological order. The corresponding lattice Lagrangian construction is explained in Appendix~\ref{app_L_to_H}.

\begin{figure}[t]
\centering
\includegraphics[width=.45\textwidth]{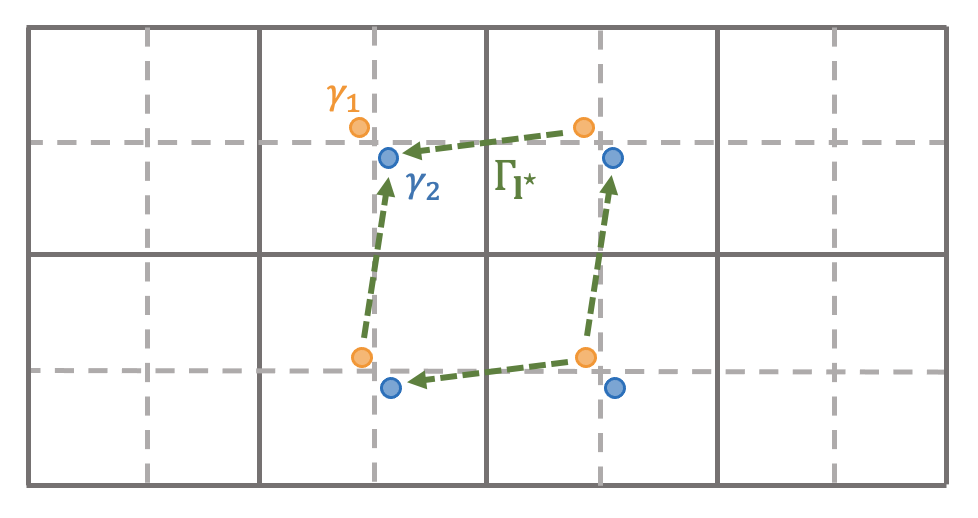}
\caption{Illustration for the $\Gamma_{\l^\star}$ operators. The arrows represent the order of the fermion operators in the product.}
\label{fig_Gamma}
\end{figure}

The way to cure the problem is to introduce to each plaquette $\p$ a fermion mode $\psi_\p, \psi_\p^\dagger$, which we recombine into two Majorana modes $\gamma^1_\p= (\psi_\p+\psi^\dagger_\p)$ and $\gamma^2_\p = \mathrm{i}(\psi^\dagger_\p-\psi_\p)$ for convenience. We do not want the low energy Hilbert space to be be expanded due to this introduction of new degrees of freedom. Instead, at low energies the fermion state should be tied to the $s^b$ boson configuration. In particular, we want the fermion number $\psi^\dagger_\p \psi_\p$ at $\p$ to be equal to the boson number $s^b_\p \mod 2$. This can be enforced energetically by adding to the Hamiltonian a term proportional to a fermion parity projector 
\begin{align}
% P &= \sum_\p \left[ \frac{1}{2} -(-1)^{s^b_\p} \left(\psi^\dagger_\p \psi_\p - \frac{1}{2}\right) \right] \nonumber \\
P &= \sum_\p \frac{ 1 + (-1)^{\psi^\dagger_\p \psi_\p + s^b_\p}}{2} \nonumber \\
&= \sum_\p \frac{1-(-1)^{s^b_\p} \, \mathrm{i}\gamma^1_\p\gamma^2_\p}{2} \ .
% P = \sum_\p \left[ \frac{1+(-1)^{s^b_\p}}{2} -(-1)^{s^b_\p} \psi^\dagger_\p \psi_\p \right] \ .
% P &= \sum_\p \left[1-\left(\psi^\dagger_\p\psi_\p-\varsigma^b_\p \right)^2\right]\nonumber\\
% &=  \sum_\p \left(1-\varsigma^b_\p \right)+ \left(2\varsigma^b_\p-1 \right)\psi^\dagger_\p\psi_\p.
\end{align} 
Moreover, we need to modify the $U_a$ term on each dual link, by multiplying to it an extra Hermitian operator
\begin{align}\label{Gamma}
   \Gamma_{\l^\star} \equiv
        \mathrm{i} (-1)^{\Sigma_{\l^\star}} \gamma^1_{\l^\star+\hat{e}_{\l^\star}/2}\gamma^2_{\l^\star-\hat{e}_{\l^\star}/2}
\end{align}
where $\hat{e}_{\l^\star} = -\hat{\mathbf{x}}$ or $\hat{\mathbf{y}}$ is the unit vector of the direction of the dual link $\l^\star$. See Fig.~\ref{fig_Gamma} for illustration. The factor $(-1)^{\Sigma_{\l^\star}}=\pm 1$ is an arbitrary background data, the lattice spin structure data, that we will explain soon. It is straightforward to verify that two such operators anti-commute if the two $\l^\star$ are separated by $\rn$, and commute otherwise. Thus, the new $U_a$ term on each link will recover the mutual commutation relation. Meanwhile, they keep the local fermion parity constraint intact, as the new $U_a$ terms commute with $P$. 

We shall think of $\Gamma_{\l^\star}$ as a Majorana hopping term across the dual link $\l^\star$. On the other hand, the $e^{\mathrm{i}d^\star\theta^a}$ part of $g^a_{\l^\star}$ involves a hopping of the $s^b$ boson. The two operators being multiplied together in the $U_a$ term means whenever an $s^b$ boson hops, a Majorana fermion hops together with it, i.e. they bind together in a $\mod 2$ sense. This explains in a physical sense why the new $U_a$ terms commute with the fermion parity projector $P$. We would like to remark that the fermion operators in our Hamiltonian only appear quadratically, in a manner that can be interpreted as hopping (in $\Gamma$) and density (in $P$); this is considerably simpler and much more intuitive than in the usual commuting projector Hamiltonians for fermion topological orders \cite{Gu:2013gma, Gaiotto:2015zta, Tarantino:2016qfy, Bhardwaj:2016clt}, which involve high powers of fermion operators in a technical manner.

The factor $(-1)^{\Sigma_{\l^\star}}$ contributes a sign that, together with the ordering of the fermion operators, determines the local Berry phase (see Appendix \ref{app_L_to_H}) picked up when the fermions hop around. But more importantly, $(-1)^{\Sigma_{\l^\star}}$ also encodes the fermion boundary conditions globally; in fact, it is a lattice version of spin structure data \cite{Gaiotto:2015zta, Tarantino:2016qfy, Bhardwaj:2016clt}. To see this, consider changing this data by $\Sigma_{\l^\star}\rightarrow \Sigma_{\l^\star} + \Xi_{\l^\star} \mod 2$ such that $(\d^\star \Xi)_{\p^\star} = 0 \mod 2$ on every dual plaquette. This does not change the Berry phase when a fermion hops around a local, contractible loop. However, when the spatial lattice has a non-trivial topology, different topological classes of $\Xi$, classified by the cohomology class $H^1(\mathbb{Z}_2)$, lead to different changes in the fermion boundary conditions, that would be witnessed if a fermion hops around non-contractible loops. (Note that the topological class of the spin structure data $\Sigma$ has no \emph{canonical} identification with $H^1(\mathbb{Z}_2)$, only the difference $\Xi$ has.)

Now that we have both the $U(1)$ background $A$ and the spin structure background $\Sigma$, we can discuss the more interesting spin-c structure on the lattice. Recall that, in the continuum in (2+1) dimensions, a spin-c background can always be regarded as a $U(1)$ background together with a spin structure, modulo the equivalence relation that any change of the spin structure can be compensated for by a suitable change of flat $\pi$ holonomy of the $U(1)$ background. Now inspect our lattice Hamiltonian. It has such an equivalence precisely when the spin-charge condition is satisfied: 
\begin{align}
    &\Sigma_{\l^\star} \rightarrow \Sigma_{\l^\star} + \Xi_{\l^\star}, \ A_\l \rightarrow A_\l+\pi\Xi_{\l-\rn} \mbox{ leaves $H$ invariant} \nonumber \\
    &\mbox{and }  \psi^\dagger_\p \psi_\p  = \rho_{\p-\rn} \mod 2\nonumber\\
    &\mbox{if and only if } \ q=0 \mod 2, \ p=k\mod 2.
    \label{spin-c_equiv}
\end{align}
That is, the Hamiltonian can be defined on a spin-c background if and only if the $U(1)$ enrichment determined by $q, p$ satisfies the spin-charge condition. Moreover, there is a remarkable difference between the equivalence relation in the continuum and the equivalence relation on the lattice. In the continuum such equivalence only makes sense globally since spin structure is a topological concept; the corresponding statement of being ``global'' on the lattice would be the aforementioned condition $(\d^\star \Xi)_{\p^\star} = 0 \mod 2$. However, the manifest equivalence \eqref{spin-c_equiv}, in fact, does not require this condition, any $\mathbb{Z}_2$ valued $\Xi$ would work. (In fact, it is also commonly required in addition that $\d^\star\Sigma$ is a suitable representative of the second Stiefel-Whitney class, chosen by the branching structure \cite{Gaiotto:2015zta, Bhardwaj:2016clt}. But now, since $\Xi$ becomes arbitrary, upon such transformation, $\d^\star\Sigma$ also becomes arbitrary.) In other words, we can conclude that \emph{the notion of spin-c background in (2+1)d lattice Hamiltonian theory is such that, the local $\mathbb{Z}_2$ spin structure data $\Sigma$ on each individual (dual) link can be completely absorbed as a $\pi$ shift into the $U(1)$ background $A$.}

\section{Solution of the Hamiltonian}
\label{sect_solution}

To solve the Hamiltonian, we adopt a ``divide and conquer'' strategy: we approximately solve one part of the Hamiltonian at a time, use the approximate solutions to project the next part of the Hamiltonian, and repeat. This works because the terms in the Hamiltonian are made ``almost commuting'' by construction, and the parameters are made to ensure sufficient separation of energy scales by assuming
% \begin{align}
%     \epsilon_a,\epsilon_b &\ll U_a,U_b, \nonumber \\
%     k^2 U_a &\ll n^2 U_b, \nonumber \\
%     V_a, V_b &\ll \sqrt{\epsilon_b U_a}, \sqrt{\epsilon_a U_b} \ .
% \end{align}
\begin{align}
    \epsilon_a,\epsilon_b,\tilde{\epsilon} \ll V_a, V_b &\ll \sqrt{\epsilon_b U_a}, \sqrt{\epsilon_a U_b} \ .
\end{align}
where $\tilde{\epsilon}\equiv \frac{k}{n} U_a \left(\frac{\epsilon_b\epsilon_a}{U_aU_b}\right)^{1/4}$. We will sketch the solution procedure in the main text; the details can be found in Appendix~\ref{app:details_solution}. 

Regarding the level of rigor of our ``controlled'' solution, we would claim the following. At each step of our procedure, we encounter local errors between our local trial solution and the actual solution. The local errors are controlled by the limits of the parameters above, but as usual, when the system size goes to the thermodynamic limit, the local errors accumulate so that the overlap between the trial solution and the actual solution tends to zero. Mathematically this may be alerting. But a crucial assumption underlying essentially all theoretical practices in condensed matter physics is that this issue does not matter -- as long as the local errors are controlled, reliable calculations can be made and physical properties can be understood. Indeed, for instance, to understand a solid-state system, instead of insisting on understanding the full microscopic Hamiltonian in one shot, we always model and study the system by starting with the low-energy electron orbits on each individual atom, and then include electron hopping and other interactions, with controlled local approximations made in the process, as long as each step has controlled local error. Similar when we 
model cold atoms in laser traps. In practice, there is no worry that the accumulation of local errors will make any theoretical modeling useless. Our solving procedure controls local error and follows the same line of reasoning, and leads to a controlled solution in the same sense.

\subsection{Bosonic theories}
\label{sect_bosonic}

We solve the low energy sectors for the bosonic cases first. The procedure to solve the fermionic cases will be analogous.

First, we solve $H^{(1)} = \sum_{\l} H^{(1)}_{\l}$ with
\begin{align}
    &H^{(1)}_{\l} =  \frac{\epsilon_a}{2} \tilde{a}_{\l^\star}^2 - U_b \cos(\d\theta^b - nb + qA)_\l
\end{align}
on each link. [Throughout this paper, if an object specified with original lattice position and an object specified with dual lattice position appear together, their positions should be understood as identical, for example, here the link $\l$ and dual link $\l^\star$ should be understood as centered at the same position, $\l=\l^\star$.] Regarding the $b$ variable as position and $\tilde{a}$ as momentum, $H^{(1)}_\l$ on each link can be thought of as a particle moving in a sinusoidal potential. When the potential minima are deep and narrow enough, there is a ``ground state'' wavefunction tightly bound to the vicinity of each potential minimum. (Of course, the true ground state is a linear combination of all of them, but the splitting is exponentially small in parameters; we will comment on this below.) We use an integer $j^b$ to label the potential minimum located at
\begin{align}
\bar{b}_\l(j^b_\l) \equiv \frac{2\pi}{n} \left( j^b +\frac{\d\theta^b+qA}{2\pi} \right)_\l,
\end{align}
and hence also labeling the (nearly degenerate) ground state wavefunction bounded to this minimum. For later convenience, we will write $j^b_\l=nz^b_\l+m^b_\l$ for $z^b_\l\in\mathbb{Z}$ and $m^b_\l\in\{0, 1, \cdots, n-1\}$. It will become useful in later steps to think of $m^b$ as an emergent $\mathbb{Z}/n\mathbb{Z}$ gauge field reduced from Higgsing the $\mathbb{R}/2\pi\mathbb{Z}$ gauge field $b$, up to a relative rescaling of $n/2\pi$. The $\mod n\mathbb{Z}$ part is parametrized by $z^b$; we will say more about this later.

Note that while $\theta^b_\v$ and $A_\l$ are $U(1)$ variables that have $2\pi$ periodicity, in the solution, a $2\pi$ shift of either amounts to a shift of the label $j^b_\l = nz^b_\l+m_\l$ by an integer. For definiteness we may fix the range of $\theta^b_\v$ and $A_\l$ to be $(0, 2\pi]$ in the below, and we will make further comments on their $2\pi$ periodicity later.

When considering $H^{(1)}_\l$ over all the lattice links together, the infinite (nearly degenerate) ground states can be labeled by the sets $\{z^b_\l\}$ and $\{m^b_\l\}$, and approximated by Gaussians (normalization factors are omitted for simplicity throughout the main text):
\begin{align} \label{firstline_solution}
    &|\{z^b_\l\},\{m^b_\l\};\{s^b_\p\},\{\theta^b_\v\}\rangle 
     \equiv \int_{\{b_\l\}} \left( \prod_\l \mathrm{e}^{-(b_\l-\bar{b}_\l)^2 /2W_b^2}\right) \bigotimes_{\l,  \p, \v}  |b_\l\rangle |s^b_\p\rangle  | \theta^b_\v \rangle \ .
\end{align}
where the Gaussian width $W_b \equiv \frac{\sqrt{2\pi}}{n} \left(\frac{\epsilon_a}{U_b}\right)^{1/4}$.
The semi-column on the left-hand-side of \eqref{firstline_solution} is used in the following way: the labels after the semi-column, such as $\{s^b_\p\},\{\theta^b_\v\}$ here, are primitive labels of the full Hilbert space regardless of the Hamiltonian, while those before the semi-column, such as $\{z^b_\l\},\{m^b_\l\}$ here, are emergent labels that only appear in low energy solutions.

The ground states has excitation gap $\Delta^{(1)} = 2\pi \sqrt{U_b \epsilon_a}$ to the first excitation states, and the trial wavefunctions in \eqref{firstline_solution} on each link span a subspace that is close to the actual ground state subspace within an error of $\mathcal{O}(\sqrt{\epsilon_a/U_b})$, which can be controlled when $\epsilon_a\ll U_b$.  Therefore, we may safely use \eqref{firstline_solution} as proxies of the ground states of the $H^{(1)}$ part of the Hamiltonian. In this limit, \eqref{firstline_solution} have very sharp, delta-function-like profile in the $b$ basis. One may note that neighboring basis states will have an overlap (tunneling) that may lead to splitting of the ground states, but that is exponentially small in $2\pi \sqrt{U_b/\epsilon_a}$ and is always less relevant than other errors in the said limits.

Next, we project $H^{(2)} = \sum_{\l^\star} H^{(2)}_{\l^\star}$ with
\begin{align}
    &H^{(2)}_{\l^\star} = \frac{\epsilon_b}{2} b_\l^2  - U_a \cos\left[(\d^\star\theta^a - n\tilde{a} + pA^\star)_{\l^\star} - \frac{k}{2}\eta_{\l^\star}b_{\l^\star+\rn}\right] 
\end{align}
onto the ground state subspace of $H^{(1)}$. [Recall $\rn \equiv \hat{\mathbf{x}}/2+\hat{\mathbf{y}}/2$, and $\eta_{\l^\star}=\mp 1$ for dual link $\l^\star$ oriented in $-\hat{\mathbf{x}}$ or $\hat{\mathbf{y}}$ direction, see Fig.~\ref{fig_lattice}; $A^\star$ is identified with $A$ through relation $A^\star_{\l^\star} = \eta_{\l^\star} A_{\l^\star+\rn}$, see Fig.~\ref{fig_lattice_A}.] After this projection, each of the two terms of $H^{(2)}_{\l^\star}$ serves an intuitive purpose. The $U_a$ term essentially shifts the integer values $\{z^b_\l\}$ and $\{s^b_\p\}$ in $|\{z^b_\l\},\{m^b_\l\};\{s^b_\p\}, \{\theta^b_\v\}\rangle$ at the same time, in such a way that keeps $(s^b-\d z^b)_\p$ on each plaquette invariant. As it will become clearer, this is because the $U_a$ term is made of $g^a_{\l^\star}$, which is the generator for the first line of \eqref{1-form_Z}. The $\epsilon_b$ term provides soft energy bound so that large shifts of $b_\l$ -- which appears in the form of large shifts of $z^b_\l$ -- are suppressed.

In order to solve the projected $H^{(2)}$, as an intermediate step we introduce a variable $\mathsf{a}_{\l^\star} \in [0, 2\pi/n)$ that is dual to the $z^b_\l$ part of $\bar{b}_\l$ (in comparison to $\tilde{a}_{\l^\star}$ which is the dual of $b_\l$):
\begin{align}\label{underline_a_definition}
    |\{\mathsf{a}_\l\},\{m^b_\l\};\{s^b_\p\}, \{\theta^b_\v\}\rangle  \equiv & \sum_{\{z^b_\l\}} \mathrm{e}^{\mathrm{i}\frac{n}{2\pi}\sum_\l \bar{b}_{\l} \cdot \mathsf{a}_{\l^\star} } |\{z^b_\l\},\{m^b_\l\};\{(s^b+\d z^b)_\p\},\{\theta^b_\v\}\rangle
\end{align}
On each $\l$, one may think of $\mathsf{a}_{\l^\star}$ as a ``quasi-momentum'' for the emergent ``lattice'' formed by ``orbitals'' $|\{z^b_\l\},\{m^b_\l\};\{s^b_\p\}, \{\theta^b_\v\}\rangle$ with different $z^b_\l$; just like quasi-momentum, the states labeled by $\mathsf{a}_{\l^\star}=0$ and $\mathsf{a}_{\l^\star}=2\pi/n$ are the same state up to an overall complex phase. In terms of this dual variable $\mathsf{a}_{\l^\star}$, the  projected $H^{(2)}_{\l^\star}$ onto the space spanned by \eqref{firstline_solution} approximately becomes:
\begin{align}\label{secondline}
   H^{(2)}_{\l^\star} & \approx \frac{\epsilon_b}{2}\left(-\mathrm{i}\frac{2\pi}{n}\partial_{\mathsf{a}_{\l^\star}}\right)^2 - U_a \cos n (\mathsf{a} - \bar{\mathsf{a}})_{\l^\star} \nonumber\\
  \bar{\mathsf{a}}_{\l^\star} & \equiv \frac{pA^\star_{\l^\star}}{n}  - \frac{ \pi k \eta_{\l^\star} }{n^2}( m^b +\frac{ \d \theta^b+qA}{2\pi})_{\l^\star+\rn}
\end{align}
Now we may change our perspective and interpret the $\mathsf{a}_{\l^\star}$ variable as a position variable on a ring. Assuming the potential is deep enough, we obtain (see Appendix~\ref{app:details_solution} for details) the approximate ground state subspace for $H^{(1)}+H^{(2)}$:
\begin{align}\label{solution_twopart}
|\{m^b_\l\} , \{s^b_\p\}; \{ \theta^b_\v \} \rangle 
\equiv & \sum_{\{z^b_\l\}} 
\ \mathrm{e}^{\i \sum_\l z^b_\l \cdot\left[ (p A^\star)_{\l^\star} - \frac{k \eta_{\l^\star}}{2n}( 2\pi m^b + \d \theta^b+qA)_{\l^\star+\rn}\right]} \nonumber\\ & \ \ \ \ \ \ \ \cdot \mathrm{e}^{-\sum_\l(\frac{n}{2\pi})^2\bar{b}_{\l}^2W_a^2/2}  |\{z^b_\l\},\{m^b_\l\};\{(s^b+\d z^b)_\p\},\{\theta^b_\v\}\rangle
\end{align} 
where the Gaussian envelope width is $1/W_a$ and $W_a \equiv \frac{\sqrt{2\pi}}{n} \left(\frac{\epsilon_b}{U_a}\right)^{1/4}$. 

We have shown in Appendix~\ref{app:details_solution} that, when $\epsilon_b\ll U_b$, $\frac{k^2}{(8n)^2} U_a\ll U_b$ and $\epsilon_a \ll U_a$, the difference between the trial wavefunctions \eqref{solution_twopart} and the actual ground states is locally bounded by $\mathcal{O}(\sqrt{\epsilon_b/U_a})$, $\mathcal{O}(\epsilon_b/U_b)$ and $\mathcal{O}(k\sqrt{U_a/U_b})$. Especially, there will be a small energy inaccuracy $\tilde{\epsilon} \sim \frac{k}{n} U_a \left(\frac{\epsilon_b\epsilon_a}{U_aU_b}\right)^{1/4}$ slightly lifting the degeneracy of different states. In the said limit, the envelope function is flat in $b$ basis since $W_a$ is small. The excitation energy due to the inclusion of $H^{(2)}$ is $\Delta^{(2)} = 2\pi \sqrt{U_a\epsilon_b}$. After this step, we created two kinds of entanglement: the $|s^b_\p\rangle$ states on the plaquettes are entangled with $|b_\l\rangle$ states on the neighboring links, and moreover, the $|b_\l\rangle$ states on links that are separated by $\rn$ are entangled together when $k\neq 0$. 

Before we proceed, we would like to make several remarks that are important for a conceptual understanding of our construction, and/or are useful for later steps.

First, we have a slight abuse of notation in \eqref{solution_twopart} -- $\{s^b_\p\}$ used to be a primitive label in the previous steps, but from here on, on the left-hand-side of \eqref{solution_twopart}, it becomes an emergent label, because what appears as a primitive label on the right-hand-side is $\{(s^b+dz^b)_\p\}$ with $\{z^b_\l\}$ summed over. The meaning of this is the following. It is useful to think of $m^b$ as an emergent $\mathbb{Z}/n\mathbb{Z}$ gauge field reduced from Higgsing the $\mathbb{R}/2\pi\mathbb{Z}$ gauge field $b$, up to a relative rescaling of $n/2\pi$, if we recall the procedure of solving for $H^{(1)}$. The $\mod n\mathbb{Z}$ part is obtained by ``gauging'' the $1$-form symmetry \eqref{1-form_Z} parameterized by $z^b$, where ``gauging'' means summing over the $z^b$ configurations, as is obtained after solving for $H^{(2)}$.

Second, now that we think of $m^b$ as an emergent $\mathbb{Z}/n\mathbb{Z}$ gauge field, an important feature is that the relation between a low energy state label by $m^b$ and a microscopic state in $b$ is not fixed, but depends on the background $A$, as can be seen from the definition of $\bar{b}$ in \eqref{firstline_solution}. This is a crucial difference compared to the usual paradigm of commuting projector Hamiltonians \cite{Levin:2004mi}. In the language of the latter, $m^b$ would be said to label a ``string type'', and a string type is tied to a fixed microscopic state on a link. In our construction they are no longer tied, allowing the construction to accommodate more possibilities, such as non-trivial Hall conductivity -- see the next remark. (If one applies the usual paradigm in the presence of Hall conductivity, as has been done in \cite{DeMarco:2021erp}, one would obtain exactly solvable lattice Hamiltonian that has discontinuities in its matrix elements, which is unphysical for microscopic lattice theory -- indeed, the construction in \cite{DeMarco:2021erp} is claimed as an effective description -- and defies the assumptions of the Kapustin-Fidkowski no-go theorem.) 

Our last remark before we proceed is, although $\theta^b_\v$ is a $U(1)$ variable, i.e. $|\theta^b_\v+2\pi \rangle = |\theta^b_\v\rangle$ by definition, in the labelling of $|\{m^b_\l\} , \{s^b_\p\}; \{ \theta^b_\v \} \rangle$ in \eqref{solution_twopart}, it is useful to regard $\theta^b_\v$ as a label that has a larger periodicity, $2\pi n$. Suppose we now let $\theta^b_\v$ take any real value on the right-hand-side of \eqref{solution_twopart}. Sending $\theta^b_\v \rightarrow \theta^b_\v+2\pi$ in \eqref{solution_twopart} does not keep the state invariant; instead, it generate a $\mathbb{Z}/n\mathbb{Z}$ gauge transformation for the $m^b_\l$ label. On the other hand, $\theta^b_\v \rightarrow \theta^b_\v+2\pi n $ preserves the state up to a phase factor. See Appendix~\ref{app:details_solution} for details. Specifically, for any set of integers $\{\tau^b_{\v}\}$,
\begin{align}
 & \ \  \left| \left\{ {m^b_\l}  \right\}, \left\{{s^b_{\p}}\right\}; \{ \theta^b_\v +2\pi n \tau^b_\v  \} \right\rangle =\mathrm{e}^{-\i \sum_{\v} \tau^b_\v \cdot \varphi_{\p^\star}} \left| \left\{ {m^b_\l}  \right\}, \left\{{s^b_{\p}}\right\}; \{ \theta^b_\v \} \right\rangle
\end{align}
with 
\begin{align}\label{bosonic_periodicphase}
    \varphi_{\p^\star} = (p \d^\star A^\star)_{\p^\star} - \frac{k }{2n}( 2\pi \d m^b +q\d A)_{\p^\star+\rn}.
\end{align}
Within the $2\pi n$ periodicity, the shift of any $\theta^b_\v$ can be regarded as the gauge transformation of $b$ field. Similarly, $A_\l$ is a $U(1)$ variable by definition, as it is equivalent to $A_\l+2\pi$ in the Hamiltonian. However, in labeling of the state \eqref{solution_twopart}, a $2\pi $ shift of $A_\l$ amounts to a change of $m^b_\l\rightarrow m^b_\l+1 \mod n$ where the $\mod n$ part will be absorbed into the neighboring $s^b_\p$. Moreover, $2\pi n$ shift of $A_\l$ will only change the neighboring $s^b_\p$ by $\pm 1$, accompanied by a phase that depends on $A$ itself -- this fact is key to having non-trivial Hall conductivity, as we will see in Sect.~\ref{sect_Hall}. In summary, while the wavefunction in primitive basis, as well as any physical result, must not depend on any $2\pi$ shift of $\theta^b_\v$ or $A_\l$ because the original Hamiltonian does not, the labeling of the basis states does. (In fact, the aforementioned discontinuity in \cite{DeMarco:2021erp} may be thought of as reminiscent of such relabeling of the states in the low energy effective description.)

Now we continue solving the remaining parts of the Hamiltonians. The remaining steps are analogous to the steps involved in solving the familiar commuting projector Hamiltonians. After obtaining the intermediate solutions \eqref{solution_twopart} for the ground states of $H^{(1)}+H^{(2)}$, we immediately find that,  up to local errors that are controllable as long as $V_b\ll\Delta^{(1,2)}$, those states are already eigenstates for $H^{(3)} = \sum_\p H^{(3)}_{\p}$ with
\begin{align}
&H^{(3)}_{\p} = \frac{V_b}{2} \left( \d b - 2\pi s^b \right)_\p^2  \ .
\end{align}
The eigenvalues are
\begin{align}
    V_b \sum_\p \left( \frac{2\pi \d m^b+q\d A}{n} - 2\pi s^b \right)^2_\p \ . \label{eigenvaluesb}
\end{align}
Thus we can identify local excitation numbers $\{v^b_\p\}$, as well as the associated energy cost from the $V_b$ (in the below $[x]$ means the nearest integer to $x$, and $[:x:]\equiv x- [x]$):
\begin{align} 
v^b_\p &\equiv -\left(\d m^b-ns^b + [\frac{q\d A}{2\pi}] \right)_\p\\
E_b &= \frac{V_b}{2} \left(\frac{2\pi}{n}\right)^2 \sum_\p \left([:\frac{q\d A}{ 2\pi}:]-v^b\right)_\p^2 \ . 
\label{excitation_b}
\end{align}
When $q(\d A)_\p/2\pi$ is half-integer the system undergoes a level crossing \cite{Levin:2011hq}. Note that each local excitation energy configuration can correspond to different sets of $\{m^b_\l\}$ and $\{s^b_\p\}$. This is because the $\mathbb{Z}/n\mathbb{Z}$ gauge transformation of $m^b$ (with the $n\mathbb{Z}_n$ part acting on $s^b$ as well) keeps the excitation configuration $\{v^b_\p\}$ invariant, and moreover, on a topologically non-trivial base manifold, there can also be different topological classes (labeled by $C$) of $\{m^b_\l\}$ and $\{s^b_\p\}$ configurations that are unrelated by $\mathbb{Z}/n\mathbb{Z}$ gauge transformation. In particular, on a torus, there are $n^2$ choices of $C$, corresponding to $n$ different closed, non-exact choices of $\mathbb{Z}/n\mathbb{Z}$ shift of $\{m^b_\l\}$ and $\{s^b_\p\}$ in each direction, that does not change $\{v^b_\p\}$.

Finally, the remaining task is to find the eigenstate of $H^{(4)} =\sum_{\p^\star}H^{(4)}_{\p^\star}$ with
\begin{align}
&H^{(4)}_{\p^\star} =  \frac{V_a}{2}\left[\left( \d^\star \tilde{a} - 2\pi s^a \right)_{\p^\star} -\frac{\pi k}{n}  s^b_{\p^\star-\rn} + \frac{k}{2n} (\d b -2\pi s^b)_{\p^\star+\rn} \right]^2 \ .
\end{align}
This should be achieved by properly superposing the intermediate solutions from the previous step, since $H^{(3)}$ and $H^{(4)}$ commute. The key point is to recognize that, in \eqref{solution_twopart}, any shift of $\{\theta^b_\v \}$  generates a gauge transformation of $b$ field in the sense that it preserves the excitation (flux) configuration of the $H^{(3)}$ terms. Therefore, it is natural to start with a set of representative $\{m^{b,rep}_{\l}\}$ and $\{s^{b,rep}_{\p}\}$ associated with an excitation configuration $\{v^{b}_\p \}$ and a topological class (which we label by $C$), and integrate all $\theta^b_\v$ over $[0,2\pi n)$ (as opposed to $[0,2\pi)$, due to the discussion above \eqref{bosonic_periodicphase}) with a carefully designated phase:
\begin{align}
& |\{v^a_{\p^\star}\}, \{v^b_\p\}, C \rangle \equiv  \int_{\{\theta^b_\v\}} \ \mathrm{e}^{-\frac{\mathrm{i}}{n} \sum_\v \theta^b_\v \cdot  \chi_{\p^\star}} \left|\{m^{b,rep}_{\l}\},  \{s^{b, rep}_\p\}; \{ \theta^b_\v  \} \right\rangle \ , \nonumber\\[.2cm]
& \chi_{\p^\star} \equiv   \left([:\frac{p\d^\star A^\star}{2\pi}:]-v^a\right)_{\p^\star} +\frac{k}{2}s^{b,rep}_{\p^\star-\rn} - \frac{k}{2n} \left([:\frac{q\d A}{2\pi}:]-v^b\right)_{\p^\star+\rn} \ .  \label{Cdefinition}
\end{align}
The phase factor designated to ensure the integrand is periodic under $\theta^b\rightarrow \theta^b+2\pi n$, and $v^a_{\p^\star}$ are arbitrary integers whose physical meaning will be identified shortly. We prove in Appendix~\ref{app:details_solution} that such a state is indeed an eigenstate for $H^{(4)}$, up to local errors that are controlled as long as $V_a \ll \Delta^{(1,2)}$. Moreover, the energy cost from the $V_a$ term for such a state can be calculated to be
\begin{align}\label{Ec}
    E_a =  \frac{V_a}{2} \left(\frac{2\pi}{n}\right)^2 \sum_{\p^\star} \left([:\frac{p\d^\star A^\star}{2\pi}:]-v^a\right)_{\p^\star}^2
\end{align}
so we immediately see that $v^a_{\p^\star}$ represents the excitation number on $\p^\star$. Thus, any such $|\{v^a_{\p^\star}\}, \{v^b_\p\}, C\rangle$ is an approximate solution to the full Hamiltonian, with energy cost $E_b+E_a$. Lastly, we note that the toric-code-like physics in $H^{(3)}$ and $H^{(4)}$ is robust to local errors resulting from the previous solving procedures of $H^{(1)}$ and $H^{(2)}$, as long as the topological excitation gaps $\sim V_a, V_b$ are large compared to those local errors of order $\epsilon_b,\epsilon_a, \tilde{\epsilon}$~\cite{Kitaev:1997wr}.

\subsection{Fermionic theories}
\label{sect_fermionic}

With the fermionic Hamiltonian, all the solving procedures for the bosonic case can be repeated with modifications that we now explain. The differences primarily come from the fact that now a plaquette state labeled by $|\{s^b_\p\}\rangle$ is always accompanied by a fermion state specified by $P$, but the sign of that fermion state remains to be specified in the first part solution \eqref{firstline_solution}. Especially, in order to solve the modified $H^{(2)}$, one still superposes different $|\{z^{b}_\l\},\{m^b_\l\};\{(s^b+\d z^{b})_\p\}, \{\theta^b_\v\} \rangle$ as in \eqref{underline_a_definition}, but their relative sign must be properly specified. 
To account for this relative fermion sign, we redefine (we will leave the fermion state as implicit in the notation below as the constraint from $P$ is always understood): 
\begin{align} \label{basis_redefinition}
    &|\{z^{b}_\l\},\{m^b_\l\};\{(s^b+\d z^{b})_\p\}, \{\theta^b_\v\} \rangle \nonumber\\ \equiv & \prod_{\l^\star} \left[(-1)^{\hat{\mathsf{z}}^{b}_{\l^\star+\rn}} \Gamma_{\l^\star} \mathrm{e}^{\mathrm{i}(\d^\star \theta^a -n \tilde{a})_{\l^\star}}\right]^{z^{b}_{\l}} |\{\mathsf{z}^b_\l=0\},\{m^b_\l\};\{s^b_\p\},\{\theta^b_\v\}\rangle
\end{align}
where the $\hat{\mathsf{z}}^b_{\l^\star+\rn}$ inside the square parenthesis should be interpreted as operators that measure the value of $\mathsf{z}^b_{\l^\star+\rn}$, so we put a hat on it to distinguish with the exponent $z^{b}_{\l^\star}$ that are merely numbers that label the state on the left-hand side. Note that different square parenthesis commute so the sequence of multiplication does not matter. Compared to the bosonic case, the newly introduced sign factors $(-1)^{\hat{\mathsf{z}}^{b}_{\l^\star+\rn}} \Gamma_{\l^\star}$ absorbs the effects of the corresponding parts in $U_a$ terms.

With this redefinition of the basis states, we obtain the same projection of the second line to the solution as in the bosonic case, \eqref{secondline}.  Therefore, repeating the solving procedure, we find that the ground state of the first two lines can still be expressed by \eqref{solution_twopart}, with the only modification of the basis states in \eqref{basis_redefinition}. 

Crucially, the ground states of the first two parts, $|\{m^b_\l\} , \{s^b_\p\} ; \{ \theta^b_\v \} \rangle$, now have different phase associated with shifting $\theta^b_\v$ by $2\pi n$, comparing with the bosonic case \eqref{bosonic_periodicphase}. The phase now becomes:
\begin{align} \label{fermionic_periodicphase}
    \varphi_{\p^\star} &= (p\d^\star A^\star+\pi k \d^\star \Sigma )_{\p^\star} - \frac{k}{2n}(2\pi \d m^b +q\d A)_{\p^\star+\rn} +\pi k (s^b_{\p^\star+\rn}+s^b_{\p^\star-\rn})
\end{align}
This leads to a generalization of the constructed bosonic solution in \eqref{Cdefinition} to fermionic cases, with the only modification on the phase factor:
\begin{align}
\chi_{\p^\star} \equiv &   \left([:\frac{\d^\star (p A^\star+k\pi \Sigma )}{2\pi}:]-v^a\right)_{\p^\star} +\frac{k}{2}s^{b,rep}_{\p^\star-\rn} - \frac{k}{2n} \left([:\frac{q\d A}{2\pi}:]-v^b\right)_{\p^\star+\rn}  \label{fermionicCdefinition}
\end{align}
and modified $V_a$ excitation energy:
\begin{align}\label{Ecmodified}
    E_a =  \frac{V_a}{2}\left(\frac{2\pi}{n}\right)^2 \sum_{\p^\star} \left([:\frac{\d^\star (pA^\star+k\pi \Sigma)}{2\pi}:]-v^a\right)_{\p^\star}^2
\end{align}
All those results clearly reduce to the bosonic case with $k$ even.

Apparently, when the spin-charge condition is satisfied, i.e. when $p$, $k$ are odd and $q$ is even, the effect of flipping $\Sigma_{\l^\star}$ between $0$ and $1$ is equivalent to shifting the corresponding $A^\star_{\l^\star}$ (which is always identified with $\eta_{\l^\star}A_{\l^\star+\rn}$) by $\pi$. This equivalence relation, necessary for the notion of spin-c, is true for all observables of the system.

\section{Wilson loop observables}
\label{sect_Wilson}

Here we examine the anyon statistics, by applying Wilson loop operators to any eigenstate $|\{v^a_{\p^\star}\}, \{v^b_\p\}, C \rangle$. Some of the calculation details can be found in Appendix~\ref{app:details_Wilson}.

The phase operator corresponding to the winding of a $v^a$-type anyon along path $\ell$ on the links is given by the Wilson loop of the $b_\l$ field: 
\begin{align}
L^b_\ell &\equiv \ \exp\left[ \i \sum_{\l \in \ell} b_\l \right]
\label{Lb_loop}
\end{align}
where the summation should be understood as the discrete analog of the oriented integral. When applied to $|\{m^{b}_{\l}\}, \{s^{b}_\p\} ; \{ \theta^b_\v  \} \rangle$, it approximates to $\exp \left[ i\frac{2\pi}{n} \sum_{\l \in \ell} \left(m^b +  \frac{\d \theta^b+qA}{2\pi} \right)_\l \right]$. When $\ell$ has two open ends, its operation on $|\{v^a_{\p^\star}\}, \{v^b_\p\}, C \rangle$ will change the $v^a_{\p^\star}$ at the two ends by $\pm 1$, which means the creation of a pair of $v^a$-type anyon and its anti-particle. Thus we can regard $L^b$ as a loop traveled by a $v^a$-type anyon. When $\ell$ is a topologically trivial loop, the Wilson loop operator won't change the eigenstate $|\{v^a_{\p^\star}\}, \{v^b_\p\}, C \rangle$ but only introduce a phase
\begin{align}
    \frac{2\pi}{n}\sum_{\p |_\ell} \left([: \frac{q \d A}{2\pi}:] - v^b \right)_\p
\end{align} 
in which the summation is over all plaquettes that are enclosed by $\ell$. This means a $v^a$-type anyon has fractional electric charge $q/n$, fractional mutual statistics $-2\pi/n$ with $v^b$-type anyon, and trivial self-statistics. When $\ell$ is a non-contractible loop, the eigenvalue phase is
\begin{align}
    \frac{2\pi}{n} \sum_{\l \in \ell} \left(m^b + \frac{qA}{2\pi} \right)_\l 
\end{align}
which can be used to identify the topological class $C$ of the state $|\{v^a_{\p^\star}\}, \{v^b_\p\}, C \rangle$.

On the other hand, the phase operator for a $v^b$-type anyon is given by the Wilson loop of the $a_{\l^\star}$ field. However, for any path $\ell^\star$ on dual links, the operator
\begin{align}
L^a_{\ell^\star} &\equiv \ \exp\left[ \i\sum_{\l^\star\in\ell^\star} a_{\l^\star} \right] \nonumber \\  &=\  \exp\left[ \i \sum_{\l^\star\in\ell^\star} \left( \tilde{a}_{\l^\star}-\frac{k \eta_{\l^\star}}{2n} b_{\l^\star-\rn} \right)\right]  
\label{La_loop}
\end{align}
involves non-commuting operators, and its meaning depends on the order of applying the non-commuting operators, with the ambiguity being a phase. For definiteness, we may consider a normal ordering protocol by moving all $\tilde{a}$ operators to the left of all $b$ operators. In Appendix~\ref{app:details_Wilson} we evaluate the effect of the operation of this operator on any $|\{v^a_{\p^\star}\}, \{v^b_\p\}, C \rangle$. When $\ell^\star$ has open ends, the operator creates a pair of $v^b$-type excitations. When $\ell^\star$ is a topologically trivial loop, the eigenvalue of the normal ordered Wilson loop is a phase 
\begin{align} \label{Laphase}
    \sum_{\p^\star|_{\ell^\star}} &\frac{2\pi}{n}\left[ \left([:\frac{\d^\star(p A^\star+\pi k \Sigma ) }{2\pi}:]-v^a\right)_{\p^\star} - \sum_\pm \frac{k}{2n} \left([:\frac{q\d A}{2\pi}:]-v^b \right)_{\p^\star\pm\rn}\right]
\end{align} 
where the summation is over dual plaquettes that are enclosed by $\ell^\star$. This shows that the $v^b$-type anyon has a fractional electric charge of $p/n-kq/n^2$ (with $p/n$ located right on $\p^\star$, and $-kq/2n^2$ on $\p^\star\pm \rn$), a fractional mutual statistics $-2\pi/n$ with $v^a$-type anyon as we saw before, and a fractional self-statistics (under exchange, not braiding) of $\pi k/n^2$. When $\ell^\star$ is a non-contractible loop, this operator generates a change to the topological class $C$ while keeping $\{v^a_{\p^\star}\}$ and $\{v^b_{\p}\}$.

\begin{figure}[t]
\centering
\includegraphics[width=.45\textwidth]{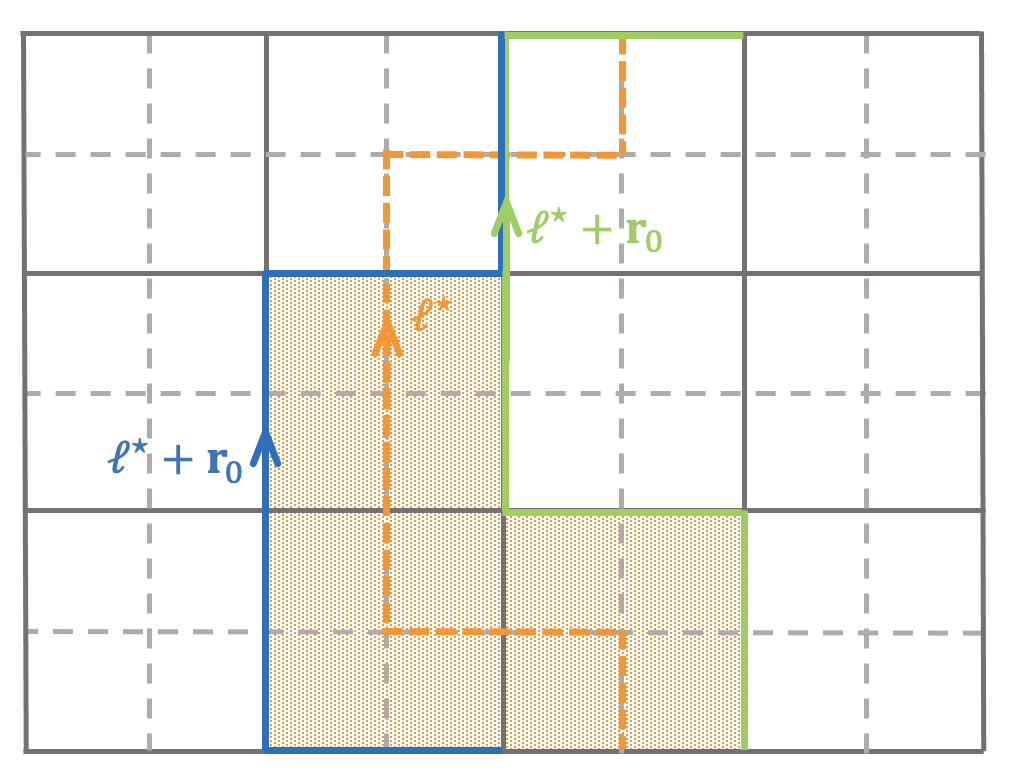}
\caption{Illustration for plaquettes (shaded) that are `sandwiched' by $\ell^\star\pm\rn$ in \eqref{La_n_phase}. }
\label{fig_Sandwich}
\end{figure}

It is interesting to study (normal ordered) $(L^a_{\ell^\star})^n$ for any non-contractible loop $\ell^\star$, since it takes $|\{v^a_{\p^\star}\}, \{v^b_\p\}, C \rangle$ back to itself, but up to a phase,
\begin{align}\label{La_n_phase}
 &\sum_{\l^\star\in\ell^\star} \left[(pA^\star+\pi k \Sigma)_{\l^\star}  -\sum_{\pm}\frac{\pi k \eta_{\l^\star}}{n}\left(m^{b, rep} +\frac{q A}{2\pi}\right)_{\l^\star\pm\rn} \right]+\pi k \sum_{\p|_{\ell^\star\pm\rn}} s^{b, rep}
\end{align}
where the summation of $\p$ is over plaquettes that are `sandwiched' between $\ell^\star$ shifted by $+\rn$ and $-\rn$, as shown in Fig.~\ref{fig_Sandwich}. Note that the spin-c equivalence relation (when $q=0\mod 2$, $p=k \mod 2$) is manifest in this non-local observable.

\section{Hall conductivity} 
\label{sect_Hall}

We now investigate the Hall response by evaluating both local charge density and the Hall conductance. 

First, we compute the expectation of the local electric charge density \eqref{charge_density}. Carrying out the calculation in Appendix \ref{app_Hallresponse}, for any low-energy eigenstate $|\{v^a_{\p^\star}\}, \{v^b_\p\}, C \rangle$ we find
\begin{align}
   \langle p s^b_\p \rangle &= \frac{p}{n}\left(v^{b} - [:\frac{q\d A}{2\pi}:]\right)_\p \ ,  \nonumber \\[.2cm]
   \langle q s^a_{\p^\star} \rangle &=  \frac{q}{n}\left(v^{a}-[:\frac{\d^\star (p A^\star+k \pi \Sigma)}{2\pi}:]\right)_{\p^\star}  - \frac{kq^2}{2n^2} \sum_{\pm} \left(v^{b}-[:\frac{q\d A}{2\pi}:] \right)_{\p^\star\pm\rn} \
\end{align}
These results have rich physical interpretations. The first property that can be understood is that, as has been seen in the previous section using a different method, the anyon excitations $v^b$ and $v^a$ have fractional electric charges (defined in \eqref{charge_density}), $p/n - kq^2/n^2$ and $q/n$ respectively; in particular, the charge of a $v^b$ excitation at $\p$ is distributed not only at $\p$ but also the nearby $\p\pm \rn$. Then we note that, in all the physical observables, including local charge and energy densities as well as Wilson loops \eqref{Lb_loop}\&\eqref{La_loop}, $v^b$ always appear in the form $\left(v^{b}-[:\frac{q\d A}{2\pi}:] \right)_\p = \left(v^{b}+[\frac{q\d A}{2\pi}]-\frac{q\d A}{2\pi} \right)_\p$, which suggests the interpretation that large $(\d A)_\p$ close to an integer multiple of $2\pi/q$ induces the same integer multiple of $v^b$ anyons, cancelling most of its own effect. Similarly one can understand the relation between $\d^\star A^\star$ and $v^a$. When $(\d A)_\p$ or $(\d^\star A^\star)_{\p^\star}$ takes half-integer multiples of $2\pi/q$ or $2\pi/p$, the system undergoes a level crossing, so that the ground state transits to a state with different anyon configuration. On the other hand, when the system is in the ``small magnetic field'' ground state with zero $v^a, v^b$, the local charge density (defined in \eqref{charge_density}) response to weak background magnetic field as (with the identification $A^\star_{\l^\star} =\eta_{\l^\star}A_{\l^\star+\rn} $ understood):
\begin{align}
    2\pi \langle \rho_\v\rangle = \left(-\frac{pq}{n}+\frac{kq^2}{2n^2}\right) \left[(\d A)_{\v-\rn} + (\d A)_{\v+\rn}\right].
\end{align}
This indeed reproduces the expected result for Hall conductivity $\sigma_H/2\pi = -2pq/n+kq^2/n^2$ since $\sigma_H\equiv \delta \rho / \delta (\d A)$. 

Below we compute the Hall conductivity as a global property, following \cite{niu1985quantized}. This computation has significant benefits over the previous local response computation. It can show that, although our solution to the Hamiltonian is approximate, given our control over the errors, the Hall conductivity must be exactly the fraction we obtain. Moreover, it depends on less details on the lattice and can be easily applied on a more general discretized space, see the next section. 

We consider a square lattice over a torus and apply a uniform background gauge field $\Vec{A}=(\alpha_x/N_x,\alpha_y/N_y)$ (where $N_i$ is the number of sites in the $i$-direction) so that there is no magnetic field, but only flat holonomies $(\alpha_x, \alpha_y)$. Now consider any ground state $|C_0(\Vec{\alpha})\rangle$ (without any anyonic excitations) in the two-parameter space where $\alpha_x$ and $\alpha_y$ vary from $0$ to $2\pi n$. Each time $\alpha_{i=x,y}$ take $2\pi\mathbb{Z}$ value, the background holonomies is gauge equivalent to the original $\vec{\alpha}=0$ case. However, while the same background is revisited $n^2$ times, the state $|C_0(\vec{\alpha})\rangle$ may not be proportional the original one $|C_0(\vec{\alpha}=0)\rangle$, until it must when both holonomies take $0$ or $2\pi n $. It is easy to see this by examining the two inequivalent non-contractible Wilson loops $L^b_{\ell_x}$, $L^b_{\ell_y}$ that characterize $C_0(\vec{\alpha})$. (When $p$,$q$ is coprime with $n$, all distinct ground states can be reached. while for other values of $p, q$ not both coprime with $n$, some of the ground states would be visited repeatedly.) One may note, however, that there is an overall phase that depends on $A$, and this is in fact related to the Chern number below. Therefore, in the two-parameter space where $\alpha_x$ and $\alpha_y$ adiabatically vary from $0$ to $2\pi n$, at the $n^2$ points where $(\alpha_x, \alpha_y)$ are gauge equivalent to $(0, 0)$, the adiabatic state has visited the degenerate ground states under the $(0, 0)$ background holonomies for $n^2$ times. The globally defined Hall conductivity is thus given by the Chern number over the $\alpha_x, \alpha_y \in [0, 2\pi n)$ space, averaged over $n^2$ visited states \cite{niu1985quantized}:
\begin{align}
   2\pi \sigma_H &= \frac{1}{n^2} \int_{0}^{2\pi n} d\alpha_x \int_{0}^{2\pi n} d\alpha_y \ \frac{\mathcal{B}}{2\pi},  \\[.3cm] 
   \mathcal{B} &\equiv -i \left( \left\langle\frac{\partial C_0}{\partial \alpha_x}\bigg{|}\frac{\partial C_0}{\partial \alpha_y} \right\rangle - \left\langle\frac{\partial C_0}{\partial \alpha_y}\bigg{|}\frac{\partial C_0}{\partial \alpha_x} \right\rangle \right) \label{Berrycurvature} \ . 
\end{align}
Here $\mathcal{B}$ is the Berry curvature in the space of holonomies, and importantly the integral of $\mathcal{B}/2\pi$ is the Chern number that must be an integer. Carrying out the calculation in Appendix \ref{app_Hallresponse}, we find the Chern number is indeed $-2pqn+kq^2$, and hence the Hall conductivity is indeed $-2pq/n+kq^2/n^2$ as expected. (In fact, in our particular model, the Berry curvature is constant over the space of holonomies, $\mathcal{B}=-pq/\pi n+kq^2/(2\pi n^2)$.) 

The key point to employing this global method is that it ensures the value of Hall conductivity we computed must be the exact value, not contaminated by the errors in our approximate solution \cite{PhysRevB.105.155130}. This is because the Chern number is an integer by definition, as long as our errors are controlled to parametrically approach zero (and hence much smaller than $1$, the spacing between possible values of Chern numbers) in the designated limit of parameters, we can be sure that within a finite range of parameters away from that limit, the Chern number must stay the same integer, and hence the Hall conductivity must stay exactly the fraction that we obtained.

\section{Generalization to triangulation}
\label{sect_latticegeneralization}
\begin{figure}[t]
\centering
\includegraphics[width=.65\linewidth]{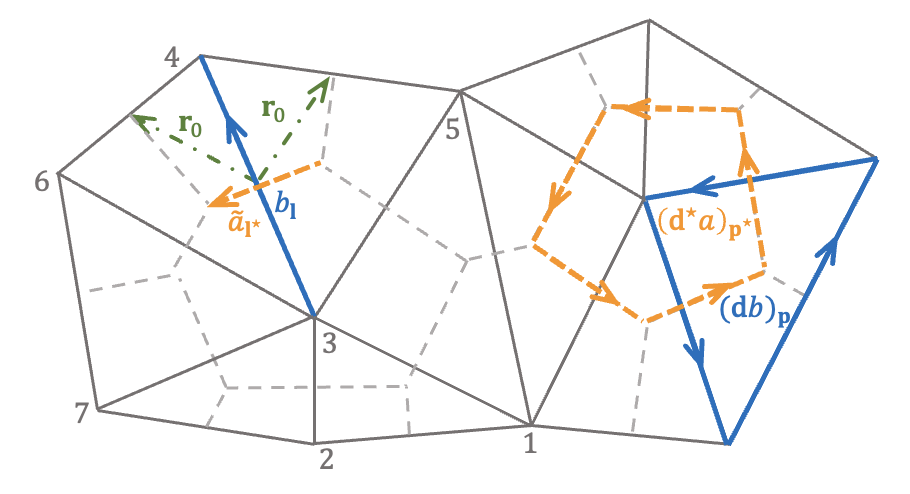}
\caption{Illustration for the triangulated lattice (solid line) and the dual lattice (dash line), the conjugate pairs of operators, and the lattice and dual lattice ``exterior derivatives''. The natural directions of the lattice links are defined to be always pointing from the vertex with a smaller index to the one with a greater index, and the natural directions of the dual lattice links are defined to be $90$-degrees counter-clockwise to the associated lattice link directions. For the example shown here, the link connecting $3,4$ has two $\rn$ pointing towards links connecting $4,5$ and $4,6$. }
\label{fig_triangulation}
\end{figure}

The constructed model can be straightforwardly generalized to any triangulated space lattice with a branching structure, which can be embedded on any base manifold. In this case, most of the equations and arguments made in our previous sections need not be modified, only some of the interpretations of the symbols need to be re-specified, which we illustrate in Fig.~\ref{fig_triangulation} and articulate as follows:
\begin{itemize}
    \item On a triangulated lattice, one needs to order all the vertices in order to give a vector field on link $\l$ a natural direction, which can be defined as the direction of the vector pointing from the end of the link with a smaller index to the other end. 
    \item The dual lattice vertices consist of the circumcenters of the triangles, and the natural directions of the dual lattice links are turned counter-clockwise compared to the associated lattice link directions.
    \item There will not be a fixed vector $\rn$. Instead, the notion of ``$\l^\star+\rn$'' will mean cup product or cap product, which is specified by the branching structure. More explicitly, consider a triangle whose three vertices are indexed as $(i,j,k)$, and $i<j<k$. Then if the link at $\l=\l^\star$ connects vertices $i$ and $j$, then `$\l^\star+\rn$' represent the {\it link} connecting $j$ and $k$. Otherwise, `$\l^\star+\rn$' is no longer meaningful for this dual link $\l^\star$. $\l^\star -\rn$ should be interpreted conversely. Importantly, there can be zero, one or two different `$\l^\star+\rn$' (or `$\l^\star-\rn$') for each $\l^\star$, since every link is shared by two triangles. An implicit summation over all possible $\rn$ should thus be understood in all the expressions containing those symbols. With the above convention for the natural directions, $\eta_{\l^\star}=1$ in the triangulated lattice.
    \item For the background gauge field, we may first define $A^\star_{\l^\star}$ on each $\l^\star$ by integrating $\vec{A}$ along this dual link, then assign the same value to all possible $A_{\l^\star+\rn}$. (In bosonic cases, alternatively, one may also define $A_\l$ on every $\l$ first, and then assign the same value to all possible $A^\star_{\l-\rn}$.)
    \item The local electric charge operator can still be defined in analogy to \eqref{charge_density}. However, it is not convenient to extract the Hall conductivity from its local response to the local magnetic field, because, first, a general triangulation is non-uniform, and the ``areas'' of the plaquettes vary, so there is no good choice of units to take a meaningful ratio between $\delta \rho$ and $\d A$, and second, this local response is not always universal anyways. The Hall conductivity computed in the global way, however, is universal and robust. It is very straightforward to generalize the global Hall conductivity calculation in Sec.~\ref{sect_Hall} to an arbitrary oriented spatial manifold with arbitrary triangulation and obtain the same result.
    \item For fermionic case, for any triangle with vertices indexed as $(i,j,k)$ with $i<j<k$, the $\Gamma$ operator on link $ij$ and that on link $jk$ should connect to the same majorana mode on the triangular plaquette; the other majorana mode, is connected by the $\Gamma$ operator on link $ik$. 
\end{itemize}

\section{Final thoughts}
\label{sect_conclusion}

We have now explained how we achieved the goals claimed in Sections \ref{sect_intro} and \ref{sect_intro2}, regarding the Hall conductivity and spin-c structure in solvable lattice Hamiltonians. The novel feature in our work compared to the established framework, as we briefly introduced at the end of Section \ref{sect_intro}, has been identified in and carried throughout the solving procedure in Section \ref{sect_solution}; we hope it may shed light on the more general developments beyond the current framework in the future.

Within the particular context of Hall conductivity in lattice models, we would like to remark that, in circumventing the Kapustin-Fikowski no-go theorem, our construction violated two of their assumptions: we have non-commuting terms in the Hamiltonian, and the local Hilbert spaces are infinite rather than finite dimensional (given these, whether the terms are projectors or not is unimportant). Since our construction is sufficient to generate a non-trivial Hall conductivity, it is natural to ask what the necessary condition is. In this regard, we would like to reiterate a conjecture and a question raised at the end of our previous work \cite{PhysRevB.105.155130}:
\begin{itemize}
    \item[$\ast$] We conjecture that the inclusion of non-commuting terms in the Hamiltonian is necessary for having Hall conductivity. If proven, this will generalize the theorem in \cite{bezrukavnikov2019localization} for non-interacting fermions to interacting systems.
    \item[$\ast$] The use of infinite dimensional local Hilbert space might not be necessary. It is interesting to see if there can be a controllably solvable model with finite dimensional local Hilbert space. However, we argue that this distinction is only interesting theoretically; in regard of physical motivations this distinction is unimportant.
\end{itemize}
See \cite{PhysRevB.105.155130} for the arguments for both.

In Section \ref{sect_intro}, a key emphasis has been the relation between the bulk lattice model construction and the gappability of the boundary. Throughout this paper, we only discussed the lattice model in the bulk, and did not discuss what happens on the gapless boundary. In future works we will carefully exam the physics on the boundary, as well as on other topological defects, in our solvable lattice models. The general theory of gapless boundaries of topological orders is an very interesting subject that is only starting to develop recently \cite{Kong:2017etd, Kong:2019byq, Kong:2019cuu}.

\acknowledgments
J.-Y.~C. is supported by NSFC under Grants No.~12174213.

\newpage

\appendix

\section{From effective Lagrangian to toy Hamiltonian}
\label{app_L_to_H}

\subsection{Bosoinic case}

In \cite{Chen:2019mjw} it was shown that the doubled Chern-Simons theory \eqref{doubleCS} has a Lagrangian description on \emph{effective} spacetime lattice -- where by \emph{effective} spacetime lattice we mean a coarse-grained spacetime manifold; the conceptual importance of this interpretation, as opposed to a ``microscopic lattice'' as in the main text, will become clear soon. For simplicity, we will think of a three-dimensional cubic lattice, but the discussions below can be straightforwardly applied to any tetrahedral decomposition of a three-dimensional manifold, or a triangulation of a two-dimensional spatial manifold together with a discretization of time (hence a ``prism'' decomposition of the spacetime). The bosonic action, with observables included, reads 
\begin{align}
S =& \ \frac{n}{2\pi} \sum_{plaq. \ p} a_p (db)_p - n\sum_{plaq. \ p} a_p s^b_p - n\sum_{link \: l} s^a_l b_l + \sum_{cube \: c} \theta^a_c (ds^b)_c + \sum_{vert. \: v} \theta^b_v (\partial s^a)_v \nonumber \\[.2cm]
& \ - \frac{q}{2\pi} \sum_{link \: l} A_l (\partial a-2\pi s^a)_l - \frac{p}{2\pi} \sum_{plaq. \: p} A^\star_p (db-2\pi s^b)_p + \sum_{plaq. \: p} a_p L^a_p + \sum_{link \: l} b_l L^b_l \nonumber\\
& \ + \frac{k}{4\pi} \sum_{link \ l} \left[ b_l(db)_{p=l+\hat{\mathbf{x}}/2+\hat{\mathbf{y}}/2+\hat{\mathbf{z}}/2} - 2\pi b_l \left( s^b_{p=l+\hat{\mathbf{x}}/2+\hat{\mathbf{y}}/2+\hat{\mathbf{z}}/2} + s^b_{p=l-\hat{\mathbf{x}}/2-\hat{\mathbf{y}}/2-\hat{\mathbf{z}}/2} \right) \right]  \ .
\label{spacetime_lattice_action}
\end{align}
Here the dynamical variables include: $a_p$ and $b_l$ which take real values on plaquette $p$ and link $l$ respectively (one may view $a_p$ as living on the dual lattice, with $l^\star=p$ in three-dimensions, and likewise for other quantities with superscript $^a$), $s^a_l$ and $s^b_p$ which take integer values on link $l$ and plaquette $p$ respectively, and $\theta^a_c$ and $\theta^b_v$ which take $U(1)=\mathbb{R}/2\pi\mathbb{Z}$ values on cube $c$ and vertex $v$ respectively. The operator $d$ is the lattice coboundary operator, i.e. lattice exterior derivative, and $\partial$ is the lattice boundary operator which is the adjoint of $d$; one may check
\begin{align}
\sum_{cube \: c} \theta^a_c (ds^b)_c &= \sum_{plaq. \: p} (\partial \theta^a)_p s^b_p, \nonumber\\
\sum_{plaq. \: p} a_p (db)_p &= \sum_{link \: l} (\partial a)_l b_l, \nonumber\\
\sum_{link \: l} s^a_l (d\theta^b)_l &= \sum_{vert. \: v} (\partial s^a)_v \theta^b_v
\end{align}
on a lattice without boundary. (In more general discretizations with branching structure, the last line of the action takes the form of cup product: $b\cup db - 2\pi s^b\cup b - 2\pi b\cup s^b$.) The background fields $A_l$ and $A^\star_p$ take real values on link $l$ and plaquette $p$ respectively; they may either independent, or identified in a certain way, say, $A^\star_p=A_{l=p+\hat{\mathbf{x}}/2+\hat{\mathbf{y}}/2+\hat{\mathbf{z}}/2}$ (in more general discretizations with branching structure, the identification is done by cap product). The $L^a_p$ and $L^b_l$, which take integer values on plaquette $p$ and link $l$ respectively, are Wilson loop observables -- they are required to be closed loops, $(\partial L^b)_v=0$, $(dL^a)_c=0$.

The action has the following gauge invariances $\!\!\mod 2\pi$:
\begin{align}
& b_l \: \rightarrow \: b_l + 2\pi z^b_l + (d\varphi^b)_l, \ \ \ \ \ s^b_p \: \rightarrow \: s^b_p + (dz^b)_p, \ \ \ \ \ \theta^b_v \: \rightarrow \: \theta^b_v + n \varphi^b_v, \\[.2cm]
& a_p \: \rightarrow \: a_p + 2\pi z^a_p + (\partial\varphi^a)_p, \ \ \ \ \ s^a_l \: \rightarrow \: s^a_l + (\partial z^a)_l, \ \ \ \ \ \theta^a_c \: \rightarrow \: \theta^a_c + n \varphi^a_c,
\label{spacetime_1-form_transf}
\end{align}
where $z^a_p$ and $z^b_l$ take integer values on plaquette $p$ and link $l$ respectively, and $\varphi^a_c$ and $\varphi^b_v$ take $U(1)=\mathbb{R}/2\pi \mathbb{Z}$ values on cube $c$ and vertex $v$ respectively. The $z^a$ and $z^b$ parameterize $1$-form $\mathbb{Z}$ gauge invariances that reduce the $a, b$ from real-valued to effectively $U(1)$ valued, and subsequently the $\varphi^a$ and $\varphi^b$ parameterize the effectively $U(1)$ ordinary ($0$-form) gauge invariances. Note the ``proper quantization'' conditions $n\in\mathbb{Z}, k\in 2\mathbb{Z}$ are used here, otherwise, the action is anomalous under the $z^a, z^b$ $1$-form gauge transformations. (In such anomalous cases, if $n$ and $k$ are some fractions, the theory will reduce to a ``smaller'' theory with new, properly quantized values $n', k'$; the $k=0$ cases were discussed in detail in the last appendix of \cite{PhysRevB.105.155130}.) 

The action also has invariances under the following changes of background variables:
\begin{align}
& A_l \: \rightarrow \: A_l + 2\pi Z_l + (d\Phi)_l, \ \ \ \ \ L^a_p \rightarrow L^a_p + q (dZ)_p, \ \ \ \ \ \theta^b_v \: \rightarrow \: \theta^b_v - q\Phi_v, \\[.2cm]
& A^\star_p \: \rightarrow \: A^\star_p + 2\pi Z^\star_p + (\partial\Phi^\star)_p, \ \ \ \ \ L^b_l \rightarrow L^b_l + p(\partial Z^\star)_l, \ \ \ \ \ \theta^a_c \: \rightarrow \: \theta^a_c - p\Phi^\star_c
\label{eff_indistinguishability}
\end{align}
where $Z_l$ and $Z^\star_p$ take integer values on link $l$ and plaquette $p$ respectively, and $\Phi_v$ and $\Phi^\star_c$ take $U(1)=\mathbb{R}/2\pi \mathbb{Z}$ values on vertex $v$ and cube $c$ respectively. Here the integrity of the electric charges $p, q$ becomes important. The $Z$ and $Z^\star$ parameterize $1$-form $\mathbb{Z}$ gauge invariances that reduce the background $A, A^\star$ from real-valued to effectively $U(1)$ valued, and subsequently the $\Phi$ and $\Phi^\star$ parametrize the effectively $U(1)$ ordinary ($0$-form) global symmetries. Apparently, if $A^\star$ is identified with $A$ in the said way, then $Z^\star, \Phi^\star$ must be identified with $Z, \Phi$ in the associated manner (but the Wilson loop observables $L^a$ and $L^b$ are still independent).

An important feature of this \emph{effective} Lagrangian theory is that the backgrounds $A_l$ and $A_l + 2\pi Z_l$ are not identical outright; an associated transformation of the Wilson loop observables is required. In physical terms, this means in this theory, a narrow thread of $2\pi$ electromagnetic flux in the background is physically indistinguishable from inserting a Wilson loop observable which creates a certain anyon worldloop. This is indeed a property of the original doubled Chern-Simons theory \eqref{doubleCS} in the continuum. A conceptually important point, however, is that this property is an effective, macroscopic one, in the sense that it only applies when the thread of the electromagnetic flux is ``narrow'' compared to the scales of interest but large compared to any microscopic scale. This is why the lattice Lagrangian theory above is only applicable to \emph{effective} spacetime lattice, by which we mean a coarse-grained spacetime manifold. Were the ``lattice'' not a coarse grain but an actual \emph{microscopic} lattice, the description is inapplicable, because a $2\pi$ magnetic flux that is narrower than the microscopic lattice scale should be invisible outright, i.e. the backgrounds $A_l$ and $A_l+2\pi Z_l$ should be identical outright. It is this important difference in the physical requirements between a theory on an effective lattice (coarse grain) and one on an actually microscopic lattice that makes the usual wisdom \cite{Levin:2004mi, Kirillov:2011mk, Bhardwaj:2016clt} which generates toy model Hamiltonians on microscopic lattice from coarse-grained descriptions (fixed point properties) not directly applicable to our present problem \cite{Chen:2019mjw}.

Let us nonetheless proceed and obtain the Hilbert space and operator contents from this effective Lagrangian. For the cubic lattice, we view two directions as the spatial lattice and one direction as the discretized time. (We may do the same for a prism decomposition of the spacetime, where the space is triangulated and the time is discretized.) Then a three-dimensional vertex $v$ is associated with a spatial vertex $\v$ and an integer time step $t$, denoted as $v=(\v, t)$, while (the center of) a three-dimensional cube $c=(\p, t+1/2)$; on the other hand, a three-dimensional link $l$ has two possibilities $l=(\l, t)$ or $l=(\v, t+1/2)$, and a three-dimensional plaquette has two possibilities $p=(\p, t)$ or $p=(\l, t+1/2)$. In these notations, our previous action reads $S = \sum_t L_t$, with
\begin{align}\label{L}
L_t \ =& \frac{1}{2\pi} \sum_\l \left(n a_{\l, t+1/2} { + \frac{k}{2} \eta_\l b_{\l-\rn,t}}\right)  \left( b_{\l, t+1} - b_{\l, t} \right)\nonumber\\ %\ \ \text{or} \ \ \frac{1}{2\pi} \sum_\l \left(n a_{\l, t+1/2} { + \frac{k}{4} \eta_\l (b_{\l-\rn,t}+b_{\l+\rn,t+1})}\right)  \left( b_{\l, t+1} - b_{\l, t} \right)\nonumber\\
&- \sum_\p \theta^a_{\p, t+1/2} \left( s^b_{\p, t+1} - s^b_{\p, t} \right)  + \sum_\v s^a_{\v, t+1/2} \left( \theta^b_{\v, t+1} - \theta^b_{\v, t} \right) \nonumber \\[.2cm]
& + \sum_\l s^a_{\l, t} \left[ d\theta^b - n b + qA \right]_{\l, t} \nonumber\\ 
& + \sum_\l s^b_{\l, t+1/2} \left\{ \left[ \partial \theta^a - n a  + pA^\star \right]_{\l, t+1/2} { -\frac{k}{2}\eta_\l b_{\l-\rn,t}-\frac{k}{2}\eta_\l b_{\l+\rn,t+1}}\right\} \nonumber \\[.2cm]
& + \sum_\p \frac{a_{\p, t}}{2\pi} \left[ n \left(d b - 2\pi s^b \right) - q dA + 2\pi L^a \right]_{\p, t} \nonumber\\ 
&+ \sum_\v \frac{b_{\v, t+1/2}}{2\pi} \left[ n \left(\partial a - 2\pi s^a \right) - p dA^\star + 2\pi L^b \right]_{\v, t+1/2} \nonumber\\
& + \sum_{\v} \frac{k b_{\v,t+1/2} }{4\pi}  \left[(db-2\pi s^b)_{\v+\rn, t+1} +(db-2\pi s^b)_{\v-\rn, t}  \right] \nonumber\\
\end{align}
where $\eta_\l=\pm 1$ when $\l$ is a link in x or y direction respectively, and $\rn \equiv \hat{\mathbf{x}}/2+\hat{\mathbf{y}}/2$. For simplicity, we have assumed that $A, A^\star$ only induce magnetic fields but no electric fields, and $L^a$, $L^b$ only run along the time direction, i.e. the anyon insertions created by inserting these Wilson loops are held at fixed positions in the space. Now we define $\tilde{a}_{\l, t+1/2} \equiv {a}_{\l, t+1/2} + (k/2n) \eta_\l b_{\l-\rn,t}$ and get
\begin{align}
L_t \ =& \ \frac{n}{2\pi} \sum_\l \tilde{a}_{\l, t+1/2} \left( b_{\l, t+1} - b_{\l, t} \right) - \sum_\p \theta^a_{\p, t+1/2} \left( s^b_{\p, t+1} - s^b_{\p, t} \right)  + \sum_\v s^a_{\v, t+1/2} \left( \theta^b_{\v, t+1} - \theta^b_{\v, t} \right) \nonumber \\[.2cm]
& \ + \sum_\l s^a_{\l, t} \left[ d\theta^b - n b + qA \right]_{\l, t} + \sum_\l s^b_{\l, t+1/2} \left\{ \left[ \partial \theta^a - n\tilde{a}  + pA^\star \right]_{\l, t+1/2}  -\frac{k}{2}\eta_\l b_{\l+\rn,t+1}\right\} \nonumber \\[.2cm]
& \ + \sum_\p \frac{a_{\p, t}}{2\pi} \left[ n \left(d b - 2\pi s^b \right) - q dA + 2\pi L^a \right]_{\p, t} \nonumber\\ 
& \ + \sum_\v \frac{b_{\v, t+1/2}}{2\pi} \left\{  \left[ n\left(\partial \tilde{a} - 2\pi s^a\right) - p dA^\star + 2\pi L^b  \right]_{\v, t+1/2} \right. \nonumber\\
    & \ \ \ \ \ \ \ \ \ \ \ \ \ \ \ \ \ \ \ \ \ \ \ \ \ \ \ \ \ \ \ \ \ \ \ \ \ \ \ \ \ \ \left. - \pi k s^b_{\v-\rn,t}+  \frac{k}{2}(db-2\pi s^b)_{\v+\rn, t+1} \right\} \ .
\end{align}
Only the three terms in the first line involve discretized time derivative; as usual, they give rise to the local Hilbert spaces endowed with the commutation relations \eqref{ba_comm}, \eqref{sathetab_comm} and \eqref{sbthetaa_comm}, as long as we introduce the spatial dual lattice notions as explained there. On the other hand, the four terms in the last two lines all involve Lagrange multipliers, so instead of Hamiltonian terms, they give rise to strict constraints on the Hilbert space. In particular, summing over the integer-valued $s^a_{\l, t}$ and $s^b_{\l, t+1/2}$ imposes the Gauss's law constraints for the $1$-form $\mathbb{Z}$ gauge transformations \eqref{1-form_Z}, generated by \eqref{generator_g}, while integrating over the real-valued $a_{\p, t}$ and $b_{\v, t+1/2}$ imposes the Gauss's law constraints for the ordinary ($0$-form) gauge transformations, generated by \eqref{generator_f}:
\begin{align}
g^b_\l = e^{-iqA_\l}, & \ \ \ \ \ g^a_{\l^\star} = e^{-ip A^\star_{\l^\star}} \ ; \\
 G^b_\p = \frac{q}{n} (\d A)_\p - \frac{2\pi}{n} L^a_{\v^\star=\p},& \ \ \ \ G^a_{\p^\star} = \frac{p}{n} (\d^\star A^\star)_{\p^\star} - \frac{2\pi}{n} L^b_{\v=\p^\star}.
\label{full_constraints}
\end{align}
The commutation relations along with these Gauss's law constraints is the full content of the theory; there is no Hamiltonian on top of the subspace specified for these constraints.

While this theory is well-defined, there are two undesired features, given that our goal is to construct microscopic lattice (toy) model Hamiltonians:
\begin{enumerate} 
\item
Again, a $2\pi$ background magnetic flux $(\d A)_\p$ through a single plaquette $\p$ is not invisible, but only equivalent to an $L^a_\p$ insertion taking value $-q$. Therefore the ``spatial lattice'' here must be viewed as a coarse grain rather than an actual microscopic spatial lattice. We are interested in theories (albeit toy theories) on the microscopic spatial lattice, in which a $2\pi$ background magnetic flux $(\d A)_\p$ through a single plaquette $\p$ is invisible outright.
\item
The current theory is a lattice gauge theory with strict Gauss's law constraints imposed on the physical Hilbert space. We are interested in theories with no strict constraints on Hilbert space, and any appearance of gauge constraint should be emergent at low energies \cite{Kitaev:1997wr, Levin:2004mi}.
\end{enumerate}
Therefore the remaining task is to modify the theory to evade these two issues.

To resolve the first issue, motivated by \cite{Levin:2011hq}, in $L_t$ we may strip off the direct coupling of the backgrounds $A, A^\star$ to the real-valued $a, b$, so that the backgrounds only couple to integer-valued variables $s^a$, $s^b$, and hence their $2\pi\mathbb{Z}$ parts indeed do not matter, and they are indeed $U(1)$. As a result of this, the $G^b, G^a$ constraints in \eqref{full_constraints} can no longer be gauge constraints, but they can emerge as energetic conditions. Consider the lattice gauge theory with two $1$-form $\mathbb{Z}$ Gauss's law constraints and a Hamiltonian:
\begin{align}
& g^b_\l = e^{-iqA_\l}, \ \ \ \ \ g^a_{\l^\star} = e^{-ip A^\star_{\l^\star}} \nonumber \\[.2cm]
& H_{gauge} = \frac{V_b}{2} \sum_\p \left(G^b_\p\right)^2 + \frac{V_a}{2} \sum_\v \left(G^a_\v\right)^2. 
\label{gauge_theory}
\end{align}
In this theory the background fields $A_\l$ and $A^\star_{\l^\star}$ are indeed $U(1)$ valued as desired. (In the above we omitted the electric field for convenience; it can be shown that the theory \eqref{gauge_theory} stays the same even if include the electric field via the time dependence of $A$ and $A^\star$.) Moreover, under the Gauss's law constraints, minimizing the Hamiltonian $H_{gauge}$ will indeed lead to the $G^a, G^b$ constraints in \eqref{full_constraints}, with $L^a_{\v^\star=\p}$ given by the integer closest to $q(\d A)_\p$ and $L^b_{\v=\p^\star}$ given by the integer closest to $p(\d^\star A^\star)_{\p^\star}$ -- this means large enough magnetic fluxes can create anyon insertions, and one reminiscence of this fact is the aforementioned macroscopic effective indistinguishably between a finite size $2\pi$ background magnetic flux and a certain Wilson loop insertion.

The theory \eqref{gauge_theory} is, however, still a gauge theory. To resolve this second issue, we recall that in the previous studies of exactly solvable models, gauge constraints are energetically imposed \cite{Kitaev:1997wr, Levin:2004mi}. This motivates us to view \eqref{gauge_theory} as the $U_a, U_b \rightarrow \infty$ limit of the ``prototype'' Hamiltonian $\wt{H}$ introduced in Section \ref{sect_Hamiltonian}. This is how we motivate for $\wt{H}$. Upon making $U_a, U_b$ any finite values, however, the local gaplessness problem arises, as explained in Section \ref{sect_Hamiltonian}. This problem occurs here but not in \cite{Kitaev:1997wr, Levin:2004mi} because the local operators here take continuous rather than discrete values, in order to accommodate for suitable couplings to the continuous background $U(1)$ gauge field(s) $A$ and $A^\star$. The resolution to this local gaplessness problem finally led us to our construction of the Hamiltonian $H$. See our previous work \cite{PhysRevB.105.155130} for a more thorough discussion in this regard (we focused on $k=0$ there but the relevant discussion does not rely on it).

\subsection{Fermionic case}

When $k$ is odd, the $1$-form gauge invariance parameterized by $z^b$ in \eqref{spacetime_1-form_transf} fails, as the path integral weight may receive a minus sign when $z^b$ is odd. In other words, the $\mathbb{Z}_2$ gauge transformation given by $z^b \mod 2$ becomes anomalous, so that the partition function always vanishes. To remedy this sign issue, we need the theory of fermionic topological orders \cite{Gu:2012ib, Gaiotto:2015zta, Bhardwaj:2016clt}.

\begin{figure}[t]
\centering
\subfigure[]{\label{fig_bloch} \includegraphics[width=.2\textwidth]{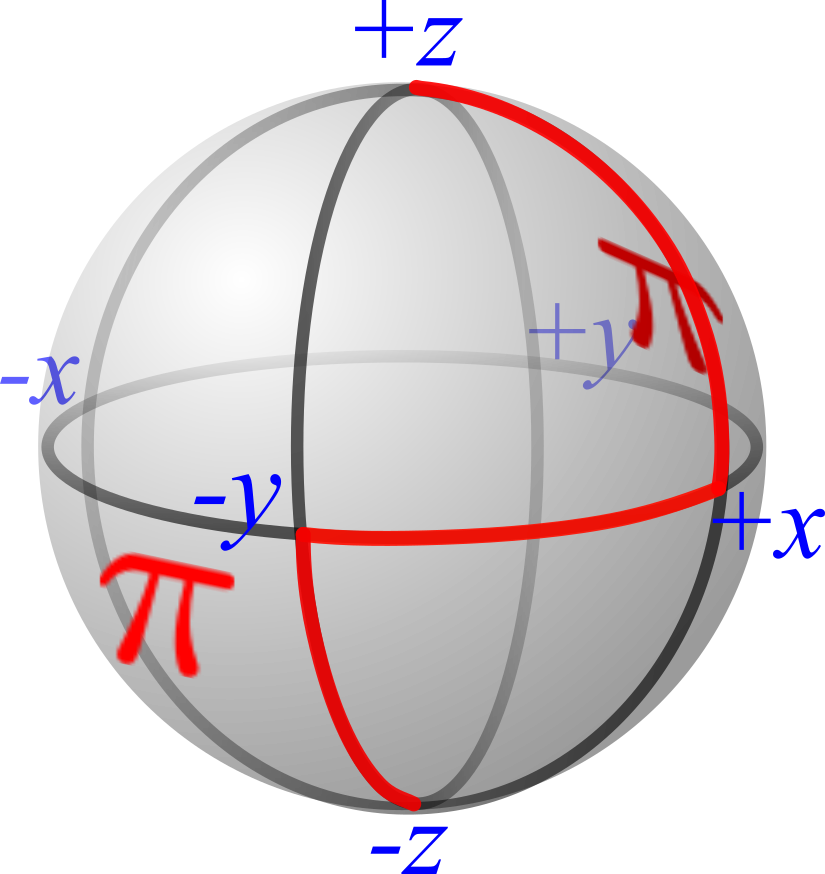}} 
\hspace{2cm}
\subfigure[]{\label{fig_abcd} \includegraphics[width=.4\textwidth]{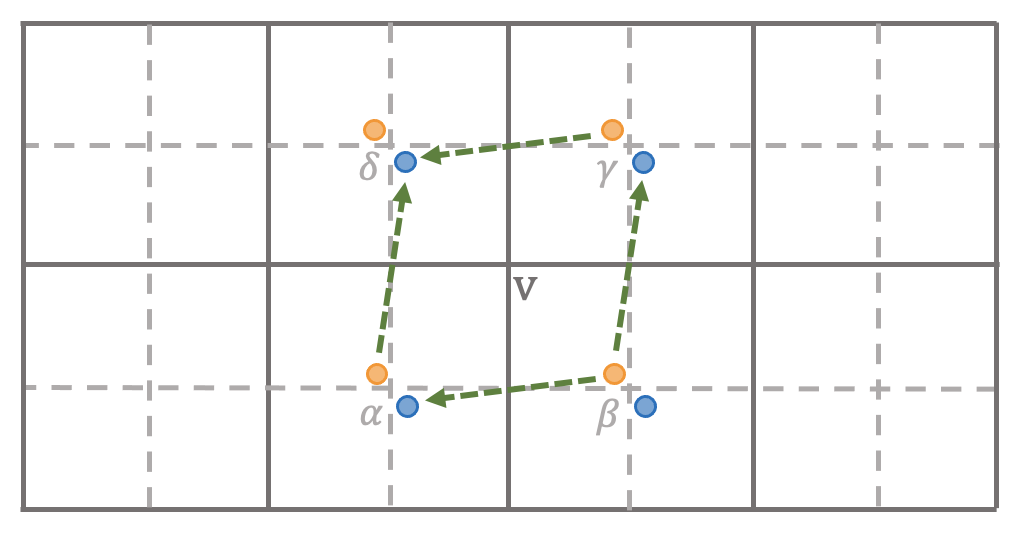}} 
\caption{(a) (Reproduced from \cite{Chen:2019mjw}.) In the Lagrangian formalism, the worldline of the fermion may travel in $\pm \hat{\mathbf{x}}$, $\pm \hat{\mathbf{y}}$ and $\pm \hat{\mathbf{t}}$ ($\pm \hat{\mathbf{z}}$) directions, indicated by the 6 discrete points on the ``Bloch sphere''. The 12 arcs connecting these points correspond to the 24 ways
that the worldline may make a turn. The 3 red arcs correspond to the 6 ways of turning that will
contribute a $(-1)$ factor to the partition function, while the remaining 9 black arcs are 18 ways of turning that
will contribute a $(+1)$ factor. We may interpret the $(\pm 1)$ factor as the exponentiation of the
integration of a Berry connection along the arc. Thus, there is a $\pi$ Berry phase around the
1/8 sphere cornered at $(+x, +y, +z)$ and the 1/8 sphere cornered at
$(-x, -y, -z)$. (b) The $\Gamma$ operator defined in the Hamiltonian formalism produces the same Berry phase. Here we also illustrate the positions of the dual vertices $\alpha, \beta, \gamma, \delta$ used in \eqref{gammaproduct1}\&\eqref{gammaproduct2}. }
\end{figure}

Lattice Lagrangians for fermionic topological orders involve a fermionic sign factor in the path integral. Upon the introduction of a local fermionic mode whose parity is tied to the $\mod 2$ reduction of some dynamical loops, this sign factor arises from the ordering of these local fermionic variables. In our odd $k$ theories, these dynamical loops are those formed by $s^b$. In \cite{Chen:2019mjw} an intuitive interpretation is given for this sign factor on the cubic lattice. It can be understood as (the exponentiation of) a Berry phase accumulated when the worldlines of those fermions make turns (with an extra $-1$ for each loop of the fermion worldline). In Fig.~\ref{fig_bloch}, we summarize the Berry connection associated with different types of turns according to \cite{Chen:2019mjw}. 

In our Hamiltonian formalism, we can show that the $\Gamma$ operator introduced in \eqref{Gamma} indeed generates such Berry phase, by examining the sign structure in $\Gamma$ operator. For example, assuming the spin structure data $\Sigma$ to be $1$,
\begin{align}
    \Gamma_{\l^\star} = (\psi^\dagger_{\l^\star+\hat{e}_{\l^\star}/2} + \psi_{\l^\star+\hat{e}_{\l^\star}/2}) (-\psi^\dagger_{\l^\star-\hat{e}_{\l^\star}/2} + \psi_{\l^\star-\hat{e}_{\l^\star}/2})
\end{align}
where the four terms correspond to different turning sequences as listed below: (recall $\hat{e}_{\l^\star} = -\hat{\mathbf{x}}$ or $+\hat{\mathbf{y}}$) 
\begin{table}[h]
    \centering
    \begin{tabular}{|c|c|c|}
    \hline
    operator & turning sequence & expected sign\\
    \hline
       $\psi^\dagger_{\l^\star\pm\hat{e}_{\l^\star}/2} \psi_{\l^\star\mp\hat{e}_{\l^\star}/2}$ & $\hat{\mathbf{t}}\rightarrow \pm \hat{e}_{\l^\star} \rightarrow \hat{\mathbf{t}} $ & $+$ (trivial) \\
       $\psi_{\l^\star+\hat{e}_{\l^\star}/2} \psi_{\l^\star-\hat{e}_{\l^\star}/2}$ & $\hat{\mathbf{t}}\rightarrow \hat{e}_{\l^\star} \rightarrow -\hat{\mathbf{t}} $ & $+$ \\
       $\psi^\dagger_{\l^\star+\hat{e}_{\l^\star}/2} \psi^\dagger_{\l^\star-\hat{e}_{\l^\star}/2}$  &  $-\hat{\mathbf{t}}\rightarrow - \hat{e}_{\l^\star} \rightarrow \hat{\mathbf{t}} $ & $-$ \\
    \hline
    \end{tabular}
\end{table}
\\
Indeed the signs in front of these terms agree with the Berry phase structure in Fig.~\ref{fig_bloch}, thus verifying the 4 independent equations out of the 12 arcs of Berry connection data in Fig.~\ref{fig_bloch}. There are 8 remaining independent equations to be verified. To do so, we need to consider the product of two neighboring $\Gamma$ operators. Consider the $\Gamma$ operators on dual links surrounding a vertex $\v$, as illustrated in Fig.~\ref{fig_abcd}, and expand the product of some pairs of operators:
\begin{align}
    \Gamma_{\v+\hat{e}_x/2} \Gamma_{\v-\hat{e}_y/2} &= (\psi^\dagger_{\gamma} + \psi_{\gamma}) (-\psi^\dagger_{\beta} + \psi_{\beta})(\psi^\dagger_{\alpha} + \psi_{\alpha}) (-\psi^\dagger_{\beta} + \psi_{\beta}) \label{gammaproduct1}\\
    \Gamma_{\v-\hat{e}_x/2}\Gamma_{\v-\hat{e}_y/2}  &=(\psi^\dagger_{\delta} + \psi_{\delta}) (-\psi^\dagger_{\alpha} + \psi_{\alpha})(\psi^\dagger_{\alpha} + \psi_{\alpha}) (-\psi^\dagger_{\beta} + \psi_{\beta}) \label{gammaproduct2}
\end{align}
then analyze some of the operator products contained and their corresponding turning sequences:
\begin{table}[h]
    \centering
    \begin{tabular}{|c|c|c|}
    \hline
    operator & turning sequence & expected sign\\
    \hline
       $\psi^\dagger_{\gamma} \psi_{\beta} \psi^\dagger_{\beta} \psi_{\alpha}$ & $\hat{\mathbf{t}}\rightarrow \hat{\mathbf{x}}\rightarrow \hat{\mathbf{y}} \rightarrow \hat{\mathbf{t}}$ & $-$\\
       $\psi_{\gamma} \psi_{\beta} \psi^\dagger_{\beta} \psi_{\alpha}$ & $\hat{\mathbf{t}}\rightarrow \hat{\mathbf{x}} \rightarrow \hat{\mathbf{y}} \rightarrow -\hat{\mathbf{t}} $ & $-$\\
       $\psi^\dagger_{\delta} \psi_{\alpha} \psi^\dagger_{\alpha} \psi_{\beta}$ & $\hat{\mathbf{t}}\rightarrow -\hat{\mathbf{x}} \rightarrow \hat{\mathbf{y}} \rightarrow  \hat{\mathbf{t}} $ & $+$\\
       $\psi_{\delta} \psi_{\alpha} \psi^\dagger_{\alpha} \psi_{\beta}$ & $\hat{\mathbf{t}}\rightarrow -\hat{\mathbf{x}} \rightarrow \hat{\mathbf{y}} \rightarrow  -\hat{\mathbf{t}}$ & $+$\\
    \hline
    \end{tabular}
\end{table}
\\
Again we verify that the signs in the product expansions match those expected ones. Also note that, when we take Hermitian conjugation for the above four operators, the corresponding worldlines will reverse, and the sign in front will remain the same. Taking into account these, we now have 8 more independent equations that are verified. Therefore we can conclude that the $\Gamma$ operator we constructed indeed endows the correct fermionic sign structure suggested by the Lagrangian approach.

\subsection{Triangulated (space) lattice}

\begin{figure}[t]
\centering
\includegraphics[width=.4\textwidth]{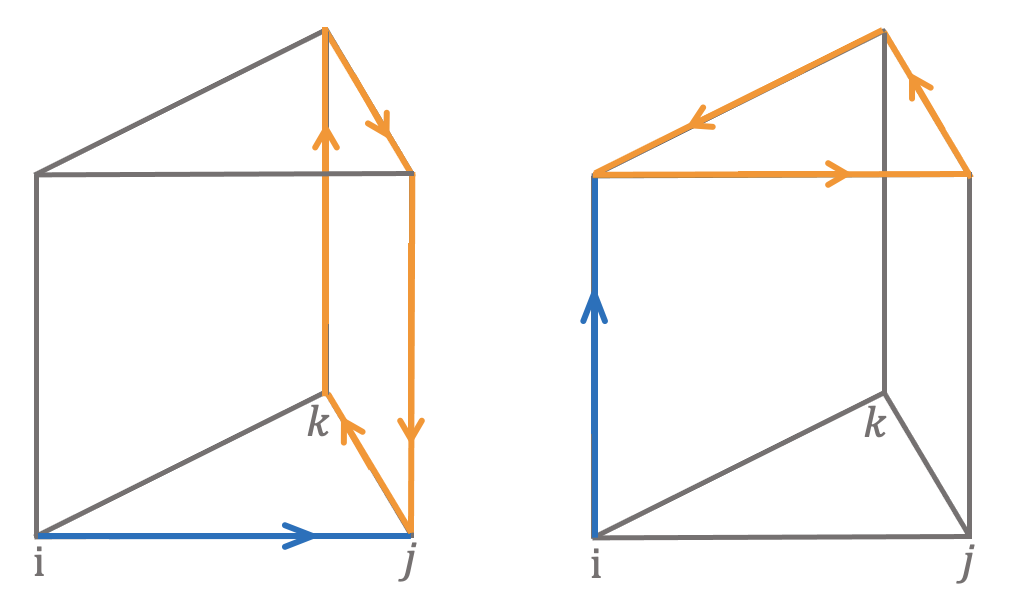}
\caption{ $b\cup db$ on each cube is a sum of two terms, assuming $i<j<k$ in the branching structure. For each term, $b$ on the blue link is multiplied to $db$ around the orange plaquette, respecting the right-hand rule. \label{fig_prism} }
\end{figure}

Here we consider triangulated space lattice, whose corresponding space-time lattice is prismatic. The Lagrangian will be the same as \eqref{L}, with the reinterpretation of $\rn$. This can be done by recognizing the meaning of $b\cup db$, see Fig.~\ref{fig_prism}.

\section{Reduction to twisted $\mathbb{Z}_n$ gauge theory in the absence of Hall conductivity}
\label{app:descrete_model}

In this appendix, we verify that the effective lattice Lagrangians we introduced above, upon ignoring the $U(1)$ global symmetry, leads to the familiar Dijkgraaf-Witten twisted $\mathbb{Z}_n$ Hamiltonians \cite{Levin:2004mi, mesaros2013classification, hu2013twisted, Lin:2014aca}. Moreover, we present another procedure that will lead to the results of \cite{PRXQuantum.3.010353}, which realizes these theories using stabilizer code. Finally, we briefly discuss the point that, with $U(1)$ global symmetry but without Hall conductivity, our effective lattice Lagrangians reproduce $U(1)$ enriched $\mathbb{Z}_n$ Hamiltonians \cite{Levin:2004mi, Wang:2021smv, wang2021exactly}.

Let's first consider the bosonic case. In the absence of the $U(1)$ global symmetry background, we sum out $s^a$ and $s^b$ in the Lagrangian \eqref{spacetime_lattice_action}. The resulting twisted $\mathbb{Z}_n$ theory reads \cite{Chen:2019mjw}:
\begin{align} \label{twisted_Zn_action}
    & Z[W,V] =  \int [D\bar{a}]^{\mathbb{Z}_n}[D\bar{b}]^{\mathbb{Z}_n}\mathrm{e}^{\mathrm{i}S[W,V]} \nonumber \\
    & S[W,V] = \int  \frac{2\pi}{n} \left[ \bar{a}\cdot\d \bar{b} - \frac{k}{2n} \bar{b}\cup \d \bar{b} \right] \ .
\end{align}
The notation $\bar{b}$ is unrelated to that used in the main text. In this appendix both $\bar{a}_l$ and $\bar{b}_l$ are $\mathbb{Z}_n$ variables. More particularly, since the $k$ twist term does not have $n\mathbb{Z}$ periodicity in $\bar{b}_l$, for definiteness we may take the range of $\bar{b}_l$ to be $\{0, \cdots, n-1\}$, and an $n\mathbb{Z}$ shift of $\bar{b}_l$ can be absorbed into a suitable $k/2$ shift of the nearby $\bar{a}_l$'s (recall $k$ is even for now). We may say the $\bar{b}$ variable is $\mathbb{Z}_n$ with a twisted periodicity.

With the same decomposition of the spacetime as in the previous section, we obtain the Lagrangian for each time slice:
\begin{align}
    L_t =& \frac{2\pi}{n} \sum_\l \left(\bar{a}_{\l, t+1/2} - \frac{k\eta_\l}{2n} \bar{b}_{\l-\rn,t}\right)  \left( \bar{b}_{\l, t+1} - \bar{b}_{\l, t}\right) + \frac{2\pi}{n}\sum_\p \bar{a}_{\p,t} (\d \bar{b})_{\p,t} \nonumber\\
    &+ \frac{2\pi}{n} \sum_\v \bar{b}_{\v, t+1/2} \left[(\partial \bar{a})_{\v,t+1/2} -\frac{k}{2n} (\d \bar{b} )_{\v+\rn,t+1} -\frac{k}{2n} (\d \bar{b})_{\v-\rn,t} \right] \ .
\end{align}
Integrating out $\bar{a}_{\p,t}$ and $\bar{b}_{\v,t+1/2}$ yields constraints that we may soften to:
\begin{align}
    L^{\text{(s)}}_t =& \frac{2\pi}{n} \sum_\l \left(\bar{a}_{\l, t+1/2} - \frac{k\eta_\l}{2n} \bar{b}_{\l-\rn,t}\right)  \left( \bar{b}_{\l, t+1} - \bar{b}_{\l, t}\right) + V_b \sum_\p \cos\frac{2\pi}{n}(\d \bar{b})_{\p,t}  \nonumber\\
    &+ V_a \sum_\v \cos \frac{2\pi}{n}\left[(\partial \bar{a})_{\v,t+1/2} - \frac{k}{2n} (\d \bar{b} )_{\v+\rn,t+1} - \frac{k}{2n} (\d \bar{b})_{\v-\rn,t}\right] \ .
\end{align}
Recognizing $|\{\bar{b}_\l\}\rangle$ as the basis that expands the Hilbert space, leads us to motivate the Hamiltonian:
\begin{align}
    H = &V_b \sum_\p \sum_{\{\bar{b}_\l\}} \cos\frac{2\pi}{n}(\d \bar{b})_\p |\{\bar{b}_\l\}\rangle \langle \{\bar{b}_\l\}| \nonumber\\ & + \frac{V_a}{2} \sum_\v \sum_{\{\bar{b}_\l\},\{\bar{b}'_\l\},\{\bar{a}_{\l^\star}\}}|\{\bar{b}'_\l\}\rangle \langle \{\bar{b}_\l\}| \nonumber\\
    & \ \ \ \ \ \ \ \ \ \ \ \ \ \ \ \ \ \ \ \ \ \ \ \ \ \ \ \ \ \cdot \mathrm{e}^{\mathrm{i} \frac{2\pi}{n} \left[(\d^\star \bar{a})_{\p^\star = \v} - \frac{ k}{2n}(\d \bar{b}')_{\v+\rn} - \frac{ k}{2n}(\d \bar{b})_{\v-\rn}+ \sum_\l \left(\bar{a}_{\l^\star=\l} - \frac{k\eta_\l}{2n} \bar{b}_{\l-\rn}\right)  \left( \bar{b}'_{\l} - \bar{b}_{\l}\right) \right]}\nonumber\\ & + \frac{V_a}{2} \sum_\v \sum_{\{\bar{b}_\l\},\{\bar{b}'_\l\},\{\bar{a}_{\l^\star}\}}|\{\bar{b}'_\l\}\rangle \langle \{\bar{b}_\l\}|\nonumber\\
    & \ \ \ \ \ \ \ \ \ \ \ \ \ \ \ \ \ \ \ \ \ \ \ \ \ \ \ \ \ \cdot \mathrm{e}^{-\mathrm{i} \frac{2\pi}{n} \left[(\d^\star \bar{a})_{\p^\star = \v} - \frac{k}{2n}(\d \bar{b}')_{\v+\rn} - \frac{k}{2n}(\d \bar{b})_{\v-\rn}-\sum_\l \left(\bar{a}_{\l^\star=\l} - \frac{k\eta_\l}{2n} \bar{b}_{\l-\rn}\right)  \left( \bar{b}'_{\l} - \bar{b}_{\l}\right) \right]} .
\end{align}
One can easily verify that this indeed corresponds to the softened Lagrangian, i.e. 
\begin{align}
    \langle\{\bar{b}_{\l,t+1}\}| \mathrm{e}^{\mathrm{i}H}|\{\bar{b}_{\l,t}\}\rangle = \mathrm{e}^{\mathrm{i}L^{\text{(s)}}_t} \ ,
\end{align} 
by appropriately plugging in the following expression of identity operator
\begin{align}
    \mathbf{1} = \sum_{\{\bar{b}_\l\},\{\bar{b}'_\l\},\{\bar{a}_{\l^\star}\}}|\{\bar{b}_\l\}\rangle \langle \{\bar{b}'_\l\}| \mathrm{e}^{\mathrm{i} \frac{2\pi}{n} \sum_{\l} \left(\bar{a}_{\l^\star=\l} { - \frac{k\eta_\l}{2n} \bar{b}_{\l-\rn}}\right)  \left( \bar{b}'_{\l} - \bar{b}_{\l}\right) } \ .
\end{align}
However, it is not obvious that this $H$ is a Hermitian operator that can serve as a Hamiltonian. This becomes apparent when we sort $H$ into a simplified form:
\begin{align}\label{discrete_H}
     H = &V_b \sum_\p \sum_{\{\bar{b}_\l\}} \cos\frac{2\pi}{n}(\d \bar{b})_\p |\{\bar{b}_\l\}\rangle \langle \{\bar{b}_\l\}| \nonumber\\ & + \frac{V_a}{2} \sum_\v \sum_{\{\bar{b}_\l\}}|\{n\lfloor:\frac{\bar{b}+\d \bar{t}^\v}{n}:\rfloor_\l\}\rangle \langle \{\bar{b}_\l\}|\mathrm{e}^{-\mathrm{i} \frac{k\pi}{n} \left[(\frac{\d \bar{b}}{n})_{\v+\rn} -  \left(\d\lfloor\frac{\bar{b}+\d \bar{t}^\v}{n}\rfloor\right)_{\v+\rn} - \sum_\l  \eta_\l \lfloor\frac{\bar{b}+\d \bar{t}^\v}{n}\rfloor_\l \bar{b}_{\l-\rn}   \right]} 
\end{align}
where $\bar{t}^\v_{\v'} = 1$ if $\v=\v'$ and $=0$ otherwise, generating quantum fluctuations between gauge equivalent configurations, and the hermitian conjugation of the second line is summed as implicit. The Hamiltonian is exactly solvable as this is a commuting Hamiltonian (the terms can be made projectors by rescaling matrix elements, but this is unimportant). 

Similarly, the fermionic twisted $\mathbb{Z}_n$ Hamiltonian for odd $k$ can be simply obtained by multiplying each term in the second line of \eqref{discrete_H} with $\prod_\l\left(\Gamma_{\l^\star=\l}\right)^{\lfloor\frac{\bar{b}+\d \bar{t}^\v}{n}\rfloor_\l}$. We omit the details of the derivation here, as it is a straightforward combination of the discussion above and the discussion in Sect.~\ref{sect_fermion}.

The Hamiltonians above, which essentially reproduce those in \cite{Levin:2004mi, mesaros2013classification, hu2013twisted, Lin:2014aca, Gu:2013gma}, might look complicated. But we may also get simpler expressions. The procedure to obtain a commuting Hamiltonian from the effective Lagrangian is not unique, and the details of the Hamiltonian depends on the procedure, though describing the same topological order. For instance, if we modify our steps above slightly, we are naturally led to a Hamiltonian that appears simpler than \eqref{discrete_H} and is identical to that found in \cite{PRXQuantum.3.010353} which emphasizes using stabilizer codes for realizing abelian topological orders.

The physical motivation of \cite{PRXQuantum.3.010353}, roughly speaking, is to make all variables explicitly periodic (without twist) when constructing Hamiltonians for twisted bosonic $\mathbb{Z}_n$ topological orders. For this purpose, instead of directly using $\mathbb{Z}_n$ with twisted periodicity as usual \cite{Levin:2004mi, mesaros2013classification, hu2013twisted, Lin:2014aca}, the construction in \cite{PRXQuantum.3.010353} is to use periodic (untwisted) $\mathbb{Z}_{n^2}$ variables and energetically condense suitable excitations in order to obtain effective $\mathbb{Z}_n$ fields with twisted periodicity. The very reason to start with an $n^2$ periodicity is, to put in our language, that the $\bar{b}_l$ variable in \eqref{twisted_Zn_action} becomes periodic if we extend their period from $n$ to $n^2$. More exactly, to obtain the $\bar{b}$ variable in \eqref{twisted_Zn_action} from the original $b$, we summed over $s^a_l, s^b_p \in \mathbb{Z}$ on every spacetime link $l$ and plaquette $p$. Now suppose we still sum over the $s^a_l$, but do the $s^b_p$ summation in two steps, the first step summing over only the $n\mathbb{Z}$ part, and the second step the remaining $\mathbb{Z}_n$ part. After the first step, the $\bar{b}_l$ variable will indeed be $\mathbb{Z}_{n^2}$, with the $n^2$ periodicity untwisted. Moreover, there would be a remaining $s^b_p \in \mathbb{Z}_n$ to be summed over, which is related to the $n\mathbb{Z}_{n^2}$ part of $\bar{b}$ via a remaining $1$-form $\mathbb{Z}_n$ gauge redundancy (think of $(d\bar{b}-n s^b)_p \mod n^2$). In obtaining \eqref{discrete_H}, we have summed over the remaining $\mathbb{Z}_n$ part of $s^b$ before turning the Lagrangian into a Hamiltonian; the reminiscent of the $\mathbb{Z}_n$ part of $s^b$ becomes $\d\lfloor\frac{\bar{b}+\d \bar{t}^\v}{n}\rfloor$. If, instead, we do not explicitly carry out the summation for the $\mathbb{Z}_n$ part of $s^b$ before turning the Lagrangian into a Hamiltonian, we will obtain a Hamiltonian identical to that in \cite{PRXQuantum.3.010353}, where the energy cost that imposes the remaining $1$-form $\mathbb{Z}_n$ redundancy corresponds to the procedure of condensing suitable excitations in \cite{PRXQuantum.3.010353}. More explicitly, the operator denoted by $C$ in \cite{PRXQuantum.3.010353} is our $1$-form gauge generator $g^a$ in \eqref{generator_g} after reducing to $\mathbb{Z}_n$, while those denoted by $A$ and $B$ in \cite{PRXQuantum.3.010353} are, respectively, exponentiations of our $G^a$ and $G^b$ in \eqref{generator_f} with coefficient $\i 2\pi/n$. Doing the same for odd $k$ gives a fermionic generalization to \cite{PRXQuantum.3.010353}.

This discussion can also be extended to cases with $U(1)$ global symmetry, as long as the enrichment is in such a manner that the Hall conductivity vanishes. In this case the lattice Lagrangian can still lead to commuting projector Hamiltonians, as allowed by the Kapustin-Fidkowski theorem. We demonstrated this for the $k=0, q=0$ untwisted cases in Section 6 of \cite{Chen:2019mjw} and Appendix A of \cite{PhysRevB.105.155130}. Roughly speaking, the idea is, similar to the above, we still explicitly sum over the $s^a$ variable in the Lagrangian, leading to a discretized $\bar{b}$ variable. But now we need to keep the $s^b$ variable before turning the Lagrangian to Hamiltonian, as $s^b$ is charged under $A$. The $\bar{b}$ variables are essentially $\mathbb{Z}_n$, with a $1$-form $n\mathbb{Z}$ gauge redundancy with $s^b$. Then, as we turn the Lagrangian to Hamiltonian, we will have reminiscents of the $G^a, G^b$ and $g^a$ energy constraints, with the latter constraint depending on $A$. This will reproduce a $U(1)$ enriched untwisted $\mathbb{Z}_n$ Hamiltonian that first appeared in \cite{Levin:2011hq}. For more general $k\neq 0$ cases without Hall conductivity, similar derivations can also be done, though we will skip them here. The resulting $U(1)$ enriched twisted $\mathbb{Z}_n$ Hamiltonians are those included in \cite{Wang:2021smv} (plus the fermionic counter-parts \cite{wang2021exactly}); it was emphasized in \cite{Wang:2021smv} that the values of $s^b$ on each plaquette that will ever be needed are actually finite, hence the local Hilbert space dimensions are all finite.

At a more abstract level, the lattice realizations for those abelian topological orders which intrinsically admit gapped boundary conditions can be summarized in the following way. In this paper, to accommodate for the most general $U(1)$ global symmetry enrichments, we need to embed the $\mathbb{Z}_n$ gauge field in the $\mathbb{Z}_n$ intrinsic topological order into the central extension sequence $2\pi\mathbb{Z} \rightarrow \mathbb{R} \rightarrow U(1)$ (up to an $n/2\pi$ rescaling between the $U(1)$ and the $\mathbb{Z}_n$) with a shifted origin. Here and below the first term in the central extension sequence is a $1$-form symmetry to be gauged. To only accommodate for those $U(1)$ global symmetry enrichments that does not have Hall conductivity, the origin does not have to be shifted and the central extension sequence can be reduced to $n\mathbb{Z} \rightarrow \mathbb{Z} \rightarrow \mathbb{Z}_n$, which is what is used in \cite{Wang:2021smv}, as explained above. When the $U(1)$ global symmetry is ignored, a general twisted intrinsic topological order uses the central extension sequence $n\mathbb{Z}_n \rightarrow \mathbb{Z}_{n^2} \rightarrow \mathbb{Z}_n$, which is what is used in \cite{PRXQuantum.3.010353}, as explained above; meanwhile an untwisted theory, i.e. the original $\mathbb{Z}_n$ toric code \cite{Kitaev:1997wr}, can use the trivial central extension sequence $1 \rightarrow \mathbb{Z}_{n} \rightarrow \mathbb{Z}_n$, which is just $\mathbb{Z}_n$ outright.

\section{Details for many-body eigenstates solution}
\label{app:details_solution}
\subsection{Solving $H^{(1)}$}
The first part of the Hamiltonian on each link is independent so that can be solved separately. Regarding $b$ variables as positions and $\tilde{a}$ variables as momenta, $H^{(1)}_\l$ on each link can be interpreted as describing a particle moving in a sinusoidal potential. Expanding the potential around one of the minima at
\begin{align}
\bar{b}_{\l,m^b_\l,z^b_\l} \equiv \frac{2\pi}{n} \left( m^b+nz^b+\frac{\d\theta^b+qA}{2\pi} \right)_\l
\end{align} 
to the quadratic order as $U_b[-1 + n^2 (b-\bar{b})^2 /2 +\dots]$, the Hamiltonian approximates to a harmonic oscillator
\begin{align}
    H^{(1)}_\l \rightarrow \omega_b (c^\dagger_{\l,m^b_\l,z^b_\l} c_{\l,m^b_\l,z^b_\l} + 1/2)
\end{align}
with excitation energy $\omega_b  = 2\pi \sqrt{U_b\epsilon_a}$, gaussian wavepacket width $W_b\equiv \frac{\sqrt{2\pi}}{n}(\frac{\epsilon_a}{U_b})^{1/4}$, and the standard construction of ladder operators $c_{\l,m^b_\l,z^b_\l} \equiv \frac{1}{\sqrt{2} W_b}[(b_\l-\bar{b}_{\l,m^b_\l,z^b_\l}) + \frac{\mathrm{i}}{n} \sqrt{\frac{\epsilon_a}{U_b}}\tilde{a}_{\l^\star}] $. The ground state solutions of the first part are thus given by \eqref{firstline_solution}.

When the gaussian wavepacket width is much less than the distance between neighboring minima,
\begin{align}\label{app:widthbcondition}
  W_b \ll \frac{2\pi}{n} \implies \frac{\epsilon_a}{U_b} \ll (2\pi)^2,  
\end{align} 
the independence of the ``harmonic oscillators'' at different minima can be justified since the overlap between neighboring orbits will be exponentially small in $(2\pi)^2/(n W_b)^2=(2\pi)\sqrt{U_b/\epsilon_a}$. Meanwhile, this condition also justifies the approximation of neglecting the higher order terms in the expansion of the potential, since those higher order terms will only introduce $\mathcal{O}(W_b^2)=\mathcal{O}(\sqrt{\frac{\epsilon_a}{U_b}})$ corrections to the low-lying states and the energy spectrum. 

\subsection{Solving $H^{(2)}$}
To solve the second part in both bosonic and fermionic cases, we project $H^{(2)}_{\l^\star}$ in the ground state subspace of the first line:
\begin{align} \label{epsilonbprojection}
    \frac{\epsilon_b}{2}b_\l^2  |\{z^b_\l\},\{m^b_\l\};\{s^b_\p\}\rangle, \{\theta^b_\v\} & \approx \frac{\epsilon_b}{2} \bar{b}_\l^2 |\{z^b_\l\},\{m^b_\l\};\{s^b_\p\},\{\theta^b_\v\}\rangle
\end{align}
\begin{align}\label{Uaprojection}
    &U_a \cos\left[(\d^\star\theta^a - n\tilde{a} + pA^\star)_{\l^\star} - \frac{k}{2}\eta_{\l^\star}b_{\l^\star+\rn}\}\right] |\{z^b_\l\},\{m^b_\l\};\{s^b_\p\}, \{\theta^b_\v\}\rangle \nonumber \\
    \approx  &\frac{U_a}{2} \sum_{\pm} \exp\pm \mathrm{i} \left[pA^\star_{\l^\star} - \frac{k}{2}\eta_{\l^\star}\bar{b}_{\l^\star+\rn} \right] |\{z^b_\l \pm \delta_{\l} \},\{m^b_\l\};\{(s^b\mp \d\delta)_\p\},\{\theta^b_\v\}\rangle\nonumber\\
    =&\frac{U_a}{2} \sum_{\pm} \exp\pm \mathrm{i} \left[pA^\star_{\l^\star} - \frac{k\eta_{\l^\star}}{2n}( 2\pi m^b + \d \theta^b+qA)_{\l^\star+\rn} \right] \nonumber \\
    & \ \ \ \ \ \ \ \ \ \ \ \ \ \ \ \ \ \ \ \ \ \ \ \ \ \ \ \ \ \ \ \ \ \ \ \ \ \ \ \ \ \ |\{z^b_\l \pm \delta^{\l^\star}_{\l} \},\{m^b_\l\};\{(s^b\pm \d\delta^{\l^\star})_\p\}, \{\theta^b_\v\}\rangle
\end{align}
where $\delta^{\l^\star}_\l = 1$ if $\l =\l^\star$ and $0$ otherwise. Note that, the derivations of the last line in bosonic and fermionic cases are due to completely different reasons: the phase $\mathrm{e}^{-\mathrm{i} \pi k \eta_{\l^\star} z^b_{\l^\star+\rn}}$ is trivial for bosonic case, while it needs to be absorbed into the definition of the basis states as in \eqref{basis_redefinition} for the fermionic case.

These terms have several crucial features that make them solvable. Firstly, they keep the $\{m^b_\l\}$ indices invariant, allowing those $\mathbb{Z}_n$ variables to remain available as labels of the ground states after solving the first two lines of the Hamiltonian. Secondly, these terms do not mix two $|\{z^b_\l\},\{m^b_\l\};\{s^b_\p\}, \{\theta^b_\v\}\rangle$ states with different $(s^b-\d z^b)_\p$ configurations. This means that the ground states of the first two lines can be obtained as a proper superposition of $|\{z^b_\l\},\{m^b_\l\};\{(s^b+\d z^b)_\p\},\{\theta^b_\v\}\rangle$ with different link states, and still be labeled by the plaquette state $\{s^b_\p\}$ These observations motivate the definition of dual variable $\mathsf{a}_{\l^\star}$ of $\bar{b}_l$ in \eqref{underline_a_definition} and the rewriting of $H^{(2)}$ in \eqref{secondline}, which we repeat here:
\begin{align}
   H^{(2)}_{\l^\star} \approx & \frac{\epsilon_b}{2}(-\mathrm{i}\frac{2\pi}{n}\partial_{\mathsf{a}_{\l^\star}})^2 - U_a \cos n (\mathsf{a} - \bar{\mathsf{a}})_{\l^\star} \ .
\end{align}
After obtaining this expression, we can again expand the potential to the quadratic order as $U_a [-1 + n^2 (\mathsf{a}-\bar{\mathsf{a}})^2+\dots]$ and approximate the Hamiltonian to that of a harmonic oscillator:
\begin{align}
    H^{(2)}_{\l^\star} \rightarrow \omega_a (d_{\l^\star}^\dagger d_{\l^\star} + 1/2)
\end{align}
with excitation energy $\omega_a  = 2\pi \sqrt{U_a\epsilon_b}$, gaussian wavepacket width $W_a\equiv \frac{\sqrt{2\pi}}{n}(\frac{\epsilon_b}{U_a})^{1/4}$, and the standard construction of ladder operators $d_{\l^\star} \equiv \frac{1}{\sqrt{2} W_a}[(\mathsf{a}_{\l^\star}-\bar{\mathsf{a}}_{\l^\star}) + \frac{\mathrm{i}}{n} \sqrt{\frac{\epsilon_b}{U_a}}\bar{b}_{\l}] $, thus obtaining the solution \eqref{firstline_solution}. Similar to the analyses in the previous subsection, the error introduced in the approximation to a harmonic oscillator can be controlled as $\mathcal{O}(W_a^2) =\mathcal{O}(\sqrt{\frac{\epsilon_b}{U_a}})$ in the limit
\begin{align}\label{app:widthacondition}
    W_a \ll \frac{2\pi}{n} \implies \frac{\epsilon_b}{U_a} \ll (2\pi)^2 \ .
\end{align}

Finally, let's justify the approximations made in \eqref{epsilonbprojection} and \eqref{Uaprojection}. In \eqref{epsilonbprojection}, the difference between the operators on the two sides can be expressed in terms of ladder operators:
\begin{align}
    \frac{1}{2}\epsilon_b (b^2 - \bar{b}^2)_\l 
    = \omega_a  \i W_bW_a\frac{n}{2\pi} (c_{\l}^\dagger + c_{\l}) (d_{\l^\star}^\dagger-d_{\l^\star}) + \omega_a  \frac{W_a^2W_b^2}{2}\left(\frac{n}{2\pi}\right)^2 (c_{\l}^\dagger + c_{\l})^2
\end{align}
where we have neglected the exponentially suppressed overlap between wavepackets centered at different $\bar{b}_\l$. Here for simplicity, we use $c_{\l}$ to represent the ladder operator that lower the energy of the state on $\l$ by $\omega_b$. It is remarkable that these terms cannot mix different states in the degenerate ground state manifold, so they can at most modify the ground states as well as the excitation gaps from the second order of perturbation. Based on this observation, we sort the terms into four groups: those $c c$, $c d$, $ c^\dagger d$, $c d^\dagger$ terms annihilate the trial ground states so that do not have effects on the ground state; the $c^\dagger c$ terms only positively modify the excitation gap $\Delta^{(1)}$, and hence unimportant; the $c^\dagger d^\dagger$ mix the trial ground states with an excited state with energy $\omega_a + \omega_b$, while the corresponding matrix element $\omega_a W_b W_a$ is already small compared to the gap, so that the mixing is controllable in the limit we already considered above. Finally, there is a $c^\dagger c^\dagger$ term with matrix element $\omega_a W_a^2W_b^2 (n/2\pi)^2/2$ which maps the ground states to an excited state with excitation energy $2\omega_b$. To justify our approximation in \eqref{epsilonbprojection}, its amplitude must be small compared to the excitation energy. This leads to another condition for the error to be controllable,
\begin{align}\label{app:energycondition}
 \omega_a \frac{W_a^2W_b^2}{2} \left(\frac{n}{2\pi}\right)^2 \ll 2\omega_b \implies \frac{\epsilon_b}{U_b} \ll (2 n)^2 \ .
\end{align}

In the second line of \eqref{Uaprojection}, the approximation occurs when we replace $b_{\l^\star+\rn}$ with $\bar{b}_{\l^\star+\rn}$. The difference between the first two lines of \eqref{Uaprojection} can be expressed as:
\begin{align}
    U_a|\cos(\frac{k}{2}b_{\l^\star+\rn} + \dots) - \cos(\frac{k}{2}\bar{b}_{\l^\star+\rn} + \dots)| \approx & U_a \frac{k}{2} (b-\bar{b})_{\l^\star+\rn} \cdot n  (\mathsf{a}-\bar{\mathsf{a}})_{\l^\star} \nonumber\\
    = &U_a \cdot [\frac{W_b}{\sqrt{2}}\frac{k}{2} (c^\dagger+c)_{\l^\star+\rn}] \cdot [\frac{W_a}{\sqrt{2}} n (d^\dagger+d)_{\l^\star}] 
\end{align}
where we have again taken into consideration that all the states that we are considering, including the low-lying excitation states, are all highly localized around $\bar{b}$'s, validating Taylor's expansion. Such terms can map the trial ground states to the excited states with energy $\omega_b+\omega_a$.  Therefore, in order to make the approximation in \eqref{Uaprojection} valid, we need to let the amplitude of this term be smaller than the excitation gap, i.e. 
\begin{align}
    U_a \frac{k n }{4}  W_a W_b \ll \omega_a + \omega_b \implies  \frac{k}{4n} U_a (\frac{\epsilon_a\epsilon_b}{U_aU_b})^{1/4} \ll \sqrt{\epsilon_a U_b} + \sqrt{\epsilon_b U_a}
\end{align}
a sufficient condition of which could be chosen as (considering $2(\epsilon_a \epsilon_b U_aU_b)^{1/4}< \sqrt{\epsilon_a U_b} + \sqrt{\epsilon_b U_a}$)
\begin{align}\label{app:Uacondition}
    \frac{k}{4n} U_a  (\frac{\epsilon_a\epsilon_b}{U_aU_b})^{1/4} \ll 2(\epsilon_a \epsilon_b U_aU_b)^{1/4} \implies \frac{U_a}{U_b} \ll (\frac{8n}{k})^2 \ .
\end{align}

Here we make a brief summary of the error analysis so far for the solution of the first two parts, \eqref{solution_twopart}. In the limits specified by \eqref{app:widthbcondition}, \eqref{app:widthacondition}, \eqref{app:energycondition}, and \eqref{app:Uacondition}, the difference of our solution to the actual ground states are bounded by errors of  $\mathcal{O}(\sqrt{\epsilon_a/U_b})$, $\mathcal{O}(\sqrt{\epsilon_b/U_a})$, $\mathcal{O}(\epsilon_b/U_b)$ and $\mathcal{O}(k\sqrt{U_a/U_b})$.

\subsection{The gauge transformations and periodicity of \eqref{solution_twopart}}

For any set of $\mathbb{Z}_n$ variables $\{t^b_{\v}\}$, we can generate gauge transformation for effective $\mathbb{Z}_n$ field in \eqref{solution_twopart} as: 
\begin{align}
|\{m^b_\l\}  , \{s^b_\p
\} , \{ \theta^b_\v +2\pi t^b_\v \}\rangle 
=&\mathrm{e}^{-\i \sum_{\l}  (\frac{m^b+\d t^b- {m^b}'}{n}) \cdot \left[ (p A^\star)_{\l^\star} - \frac{k \eta_{\l^\star} }{2n}( 2\pi {m^b}' + \d \theta^b+qA)_{\l^\star+\rn}\right]} \nonumber\\
& \ \ \ \ \ \ \ \ \ \ \ \ \ \ \ \ \ \ \ \ \ \ \ \ \ \ \ \ \ \ \ \ \ \ \ \ \ \ \  \left| \left\{ {m^b_\l}'  \right\}   , \left\{{s^b_{\p}}'\right\}, \{ \theta^b_\v \} \right\rangle \label{app:Tdefinition}
\end{align}
where
\begin{align}
    m^b_\l \rightarrow {m^b_\l}' = n\lfloor: \frac{m^b+\d t^b}{n}:\rfloor_\l  \ , \ \ s^b_{\p} \rightarrow  {s^b_{\p}}' =  \left(s^b - \d \lfloor \frac{m^b+\d t^b}{n}\rfloor\right)_\p
\end{align}
are the $\mathbb{Z}_n$ states after gauge-transformation, in which $\lfloor x\rfloor$ represents the nearest integer that is not greater than $x$, and $\lfloor: x:\rfloor=x-\lfloor x\rfloor$.

To see \eqref{solution_twopart} have different periodic boudary conditions for $\theta^b$ variables in bosonic and fermionic cases, \eqref{bosonic_periodicphase}\&\eqref{fermionic_periodicphase}, we note that the shifting of $\theta^b_\v$ by $2\pi n$ in the fermionic case play the role of applying $\mathrm{e}^{-\mathrm{i}(\d \tau^b)_\l c_{\l^\star} }$ and thus can connect different basis states, according to \eqref{basis_redefinition}:
\begin{align}
    &|\{z^{b}_\l\},\{m^b_\l\};\{(s^b+\d z^{b})_\p\},\{\theta^b_\v+2\pi n \tau^b_\v\}\rangle\nonumber\\
    \equiv& (-)^{\tau^b_{\v} \cdot\left[ (s^b+\d z^b)_{\p^\star-\rn}+s^b_{\p^\star+\rn} + (\d \Sigma)_{\p^\star}\right]} |\{(z^{b}+\d \tau^b)_\l\},\{m^b_\l\}; \{(s^b+\d z^{b})_\p\},\{\theta^b_\v\} \rangle \ .
\end{align}

\subsection{Solving $H^{(3)}$ and $H^{(4)}$}

It is straightforward to verify that the state constructed in \eqref{Cdefinition} or its fermionic generalization in \eqref{fermionicCdefinition} is diagonal for  $V_a$ term. To do so, we expand the wave function as:
\begin{align}\label{app:Cexpanded}
|\{v^a_{\p^\star}\}, \{v^b_\p\}, C \rangle \equiv & \sum_{\{z^b_\l\}}  \int_{\{\theta^b_\v\}} \mathrm{e}^{- \frac{\i}{n} \sum_{\v} \theta^b_\v \chi_{\p^\star}}\left(\prod_\l
\ \mathrm{e}^{\i z^b_\l \cdot\left[ (p A^\star)_{\l^\star} - \frac{k \eta_{\l^\star}}{2n}( 2\pi m^{b, rep} + \d\theta^b +qA)_{\l^\star+\rn}\right]} \right. \nonumber\\
&
\ \ \ \ \ \ \ \ \ \ \ \ \ \ \ \ \ \ \ \ \   \left. \mathrm{e}^{-(\frac{n}{2\pi})^2\bar{b}_{\l}^2W_a^2/2} \right) |\{z^b_\l\};\{m^{b,rep}_\l\},\{\theta^b_\v\},\{(s^{b,rep}+\d z^b )_\p\}\rangle \ .
\end{align}
We notice that when acting on a wavefunction in the basis of $|b_\l\rangle$ and $|\theta^b_\v\rangle$,
\begin{align}
    \tilde{a}_{\l^\star} = -\i \frac{2\pi}{n} \partial_{b_\l} \ , \ \ s^a_{\p^\star} = - \i \partial_{\theta^b_\v} \ .
\end{align}
Utilizing the above relation, it is straightforward to verify the following results: before summing over $\{z^b_\l \}$,
\begin{align}
    (\d^\star \tilde{a}-2\pi s^a)_{\p^\star} &\rightarrow \frac{\pi k}{n}  (\d z^b)_{\p^\star - \rn} +\frac{2\pi}{n}\chi_{\p^\star} \label{Vaapproximation} \ , \\
    -\pi k s^b_{\p^\star-\rn} &\rightarrow  - \pi k (s^{b, rep}+\d z^b)_{\p^\star-\rn} \ , \\
    \frac{k}{2} (\d b -2\pi s^b)_{\p^\star+\rn} &\rightarrow \frac{k}{2} (\frac{2\pi}{n}\d m^b +\frac{q\d A}{n} - 2\pi s^{b, rep}) _{\p^\star+\rn} \ .
\end{align}
After they sum up, the  $z^b$ dependence is canceled (hallmarking that $|C\rangle$ is an eigenstate), and we get the expression for $V_a$ excitation energy, $E_a$, in \eqref{Ec}\&\eqref{Ecmodified}.

In these steps, aside from the exponentially suppressed errors originating from the overlap between states with different $\{m^b_\l\}$ or $\{z^b_\l\}$, the main approximation occurs when we neglect the difference between $b_\l$ and $\bar{b}_\l$, and neglect the $W_a^2 (\frac{n}{2\pi})^2 (\d \bar{b})_{\p^\star}$ in \eqref{Vaapproximation}. Recovering them in $H^{(3)}$ and $H^{(4)}$:
\begin{align}
    H^{(3)}_\p &=  \frac{V_b}{2}  \left[\frac{2\pi}{n}\left([:\frac{q\d A}{ 2\pi}:]-v^b\right)+\frac{W_b}{\sqrt{2}}\d(c^\dagger+c)\right]_\p^2 \ , \\[.2cm]
    H^{(4)}_{\p^\star} & =   \frac{V_a}{2}  \left[\frac{2\pi}{n}\left([:\frac{\d^\star (pA^\star+k\pi \Sigma)}{2\pi}:]-v^a\right)_{\p^\star} + \frac{\i W_a}{\sqrt{2}}\d(d^\dagger+d)_{\p^\star} +\frac{k}{2n}\frac{W_b}{\sqrt{2}}\d(c^\dagger+c)_{\p^\star+\rn} \right]^2 \ .
\end{align}
As long as we are considering the low-lying states, the ``flux numbers'' $\left([:\frac{q\d A}{ 2\pi}:]-v^b\right)_\p$ and $\left([:\frac{\d^\star (pA^\star+k\pi \Sigma)}{2\pi}:]-v^a\right)_{\p^\star}$ are $\mathcal{O}(1)$ numbers. Under this circumstance, the errors introduced by those neglected terms are of order $W_b V_b$, $W_a V_a$ and $W_b V_b$. As long as $V_a, V_b \ll \omega_a, \omega_b$, the above errors can be safely controlled.

\section{Details for Wilson loop operators}

\label{app:details_Wilson}
In this section, we evaluate the effect of the Wilson loop operator $L^a_{\ell^\star}$ in \eqref{La_loop} that is further normal ordered, with all $c$ operators arranged to the left of $b$ operators. 

When $\ell^\star$ is a close loop and the operator is applied to state $|\{z^b_\l\};\{m^{b,rep}_\l\},\{\theta^b_\v\},\{(s^{b,rep}+\d z^b )_\p\}\rangle$, the $\mathrm{e}^{-\i\frac{k \eta_{\l^\star}}{2n} \sum_{\l^\star \in \ell^\star}b_{\l^\star-\rn}}$ part takes phase 
\begin{align}
    -\frac{k \eta_{\l^\star}}{2n} \sum_{\l^\star\in \ell^\star} b_{\l^\star-\rn} \rightarrow \  -\frac{\pi k }{n^2} \sum_{\p^\star} \left(\d m^b + n \d z^b +\frac{q\d A}{2\pi}\right)_{\p^\star-\rn} \ .
\end{align}
Meanwhile, the $\mathrm{e}^{\i \sum_{\l^\star \in \ell^\star}\tilde{a}_{\l^\star}}$ part changes $m^b_\l$ by $-1$ for all link crossed by $\ell^\star$, which is equivalent to shifting $\theta^b_{\v}$ by $+2\pi$ for all the $\v=\p^\star$ enclosed by $\ell^\star$ (assuming $\ell^\star$ is anti-clockwise oriented).
Therefore, when the operator is applied to $|C\rangle$, one can make the substitution $\theta^b_\v \rightarrow \theta^b_\v - 2\pi$ in the wavefunction of \eqref{app:Cexpanded},  then the effect of this part is introducing phase before summing over $\{z^b_\l \}$:
\begin{align}
  & \sum_{\l^\star\in \ell^\star} \tilde{a}_{\l^\star} \rightarrow  \ \frac{2\pi}{n}\sum_{\p^\star} \chi_{\p^\star} +  \frac{\pi k }{n} \sum_{\p^\star} (\d z^b)_{\p^\star-\rn} \ .
\end{align} 
The combined phase factor thus has no $z^b$ dependence, which means that the total effect of the $L^a_{\ell^\star}$ operator is purely introducing a phase, \eqref{Laphase}, to $|C\rangle$.

When $\ell^\star$ is a non-contractible loop, the operation of $L^a_{\ell^\star}$ map between different topological classes. It is of particular interest to discuss the effect of normal ordered $(L^a_{\ell^\star})^n$ since it maps $|C\rangle$ back to itself up to a phase. In this case, the effect of $\mathrm{e}^{-\i\frac{k\eta_{\l^\star}}{2} \sum_{\l^\star \in \ell^\star}b_{\l^\star-\rn}}$ on $|\{z^b_\l\};\{m^{b,rep}_\l\},\{\theta^b_\v\},\{(s^{b,rep}+\d z^b )_\p\}\rangle$ is introducing phase:
\begin{align}
    -\frac{k\eta_{\l^\star}}{2} \sum_{\l^\star\in \ell^\star} b_{\l^\star-\rn} \rightarrow \ - \frac{\pi k \eta_{\l^\star}}{n} \sum_{\l^\star\in \ell^\star} \left( m^{b,rep} + n z^b +\frac{q A}{2\pi}\right)_{\l^\star-\rn} \ .
\end{align}
Meanwhile, the effect of $\mathrm{e}^{\i n \sum_{\l^\star \in \ell^\star}\tilde{a}_{\l^\star}}$ is, roughly speaking, shifting $z^b_\l$ by $-1$ for all link crossed by $\ell^\star$ in the basis states of \eqref{solution_twopart}, which results in a phase
\begin{align}
   n \sum_{\l^\star \in \ell^\star}\tilde{a}_{\l^\star} \rightarrow \sum_{\l^\star \in \ell^\star}  (pA^\star)_{\l^\star}  - \frac{\pi k \eta_{\l^\star}}{n} \sum_{\l^\star\in \ell^\star} \left( m^{b, rep}  +\frac{q A}{2\pi}\right)_{\l^\star+\rn} \ .
\end{align}
In fermionic case, there are additional phases $\prod_{\l^\star \in \ell^\star} (-1)^{z^b_{\l^\star+\rn}+\Sigma_{\l^\star}}$, and $\prod_{\p} (-1)^{(s^{b, rep}+\d z^b)_\p}$ for plaquettes sandwiched by $\ell^\star\pm \rn$. Taking into account all the phases, one obtains the general expression of the overall phase in \eqref{La_n_phase}.

\section{Details for Hall conductivity and fractionalized electric charge}
\label{app_Hallresponse}

We first calculate the expectation value of the local electric charge density. In the presence of a background $A$ and arbitrary anyon excitations $\{v^b_\p\}$ and $\{
v^a_{\p^\star}\}$, using the expanded expression of $|\{v^b_\p\},\{
v^a_{\p^\star}\},C\rangle$ in \eqref{app:Cexpanded}, we have:
\begin{align}
   \langle  p s^b_\p \rangle 
   &= \frac{1}{\mathcal{N}}\sum_{\{z^b_{\l}\}}  \int \prod_{\v}  d\theta^b_\v \ p\left(s^{b, rep}+\d z^b\right)_\p 
   \prod_{\l} \mathrm{e}^{-\bar{b}_{\l}^2 (n/2\pi)^2 W_a^2} 
\end{align}
where we formally recovered a normalization factor $\mathcal{N}$. As far as single-link properties are concerned, the integral over $\{\theta_\v\}$ in $[0,2\pi n )$ is equivalent to a infinite integral over the $\bar{b}$ variable on every link. Therefore, the expectation can be evaluated as:
\begin{align}
   \langle   p s^b_\p   \rangle &= \frac{1}{\mathcal{N}} \prod_{\l} \int  d\bar{b}_\l \cdot \frac{p}{n}\left(\d\bar{b}+n s^{b, rep}-\d m^{b,rep} - \frac{q\d A}{2\pi}\right)_\p 
   \mathrm{e}^{-\bar{b}^2 (n/2\pi)^2 W_a^2} \nonumber
   \\ &= \frac{p}{n}\left(v^{b,C} - [:\frac{q\d A}{2\pi}:]\right)_\p \ . %=  -\frac{p}{n}\frac{q(\d A)_\p}{2\pi} + \frac{p}{n}\left(v^b + [\frac{q\d A}{2\pi}]\right) _\p.
\end{align}

On the other hand, 
\begin{align}
   \langle  q s^a_{\p^\star}  \rangle &= \frac{1}{\mathcal{N}}\sum_{\{z^b_{\l}\}}  \int \prod_{\v}  d\theta^b_\v \ \frac{q}{n}\left[-\chi_{\p^\star}-\frac{k}{2}(\d z^b)_{\p^\star-\rn}\right]
   \prod_{\l} \mathrm{e}^{-\bar{b}_{\l}^2 (n/2\pi)^2 W_a^2} \ , 
\end{align}
and we can also use the above trick to rewrite:
\begin{align}
   \langle  q s^a_{\p^\star} \rangle &= \frac{1}{\mathcal{N}}\sum_{\{z^b_{\l}\}}  \int \prod_{\v}  d\theta^b_\v \ \frac{q}{n}\left[-\chi_{\p^\star}-\frac{k}{2n}(\d \bar{b} -\d m^{b, rep} -\frac{q\d A}{2\pi})_{\p^\star-\rn}\right]
   \prod_{\l} \mathrm{e}^{-\bar{b}_{\l}^2 (n/2\pi)^2 W_a^2} \nonumber\\
   &= \frac{q}{n}\left(v^{a,C}-[:\frac{\d^\star (p A^\star+k \pi \Sigma)}{2\pi}:]\right)_{\p^\star} - \frac{kq^2}{2n^2} \sum_{\pm} \left(v^{b,C}-[:\frac{q\d A}{2\pi}:] \right)_{\p^\star\pm\rn} \ .
\end{align}

Then we calculate the Berry curvature $\mathcal{B}$ in \eqref{Berrycurvature} for the global method for Hall conductance. For simplicity, we start with the state with class $C_0$ in which $m^{b, rep}_{\l}=0$, $s^{b, rep}_{\p}=0$, $v^b_\p=0$ and $v^a_{\p^\star}=0$, so that the state is a ground state and it can be simplified to:
\begin{align}
|C_0\rangle \equiv & \sum_{\{z^b_\l\}}  \int_{\{\theta^b_\v\}} 
\left(\prod_\l
\ \mathrm{e}^{\i z^b_\l \cdot\left[ (p A^\star)_{\l^\star} - \frac{k \eta_{\l^\star}}{2n}(\d\theta^b +qA)_{\l^\star+\rn}\right]} \mathrm{e}^{-\bar{b}_{\l}^2W_a^2/2(2\pi)^2} \right) \nonumber\\
& \ \ \ \ \ \ \ \ \ \ \ \ \ \ \ \ \ \ \ \ \ \ \ \ \ \ \ \ \ \ \ \ \ \ \ \ \ \ \ \ \ \ \ |\{z^b_\l\};\{m^b_\l=0\},\{\theta^b_\v\},\{(\d z^b)_\p\}\rangle \ .
\end{align}
We note that the Berry curvature is twice the imaginary part of $\langle\frac{\partial C_0}{\partial \alpha_x}|\frac{\partial C_0}{\partial \alpha_y}\rangle$, so one of the derivative must act on the amplitude and the other must act on the phase of the wavefunction to have non-zero contribution. Because of the translational invariance of the problem, the curvature receives the same contribution from all links. Therefore, we may simply consider the contribution from one link $\l_0$, i.e. the terms that contain derivative on the amplitude through $\partial/\partial A_{\l_0}$, and then multiply this contribution by the number of links in the system. This approach yields:
\begin{align}
   \mathcal{B} &=  4\sum_{z^b_{{\l_0}}}  \int_0^{2\pi n} d\theta^b_{{\v_1}}d\theta^b_{{\v_2}} \left(-\frac{pq}{ n}+\frac{kq^2}{2n^2}\right)z^b_{\l_0}\bar{b}_{\l_0} \left(\frac{n}{2\pi}\right)^2W_a^2 \mathrm{e}^{-\bar{b}_{\l_0}^2  (n/2\pi)^2 W_a^2} / \mathcal{N}' \ , \nonumber \\[.2cm]
    \bar{b}_{\l_0} &\equiv 2\pi  \left(z^b_{\l_0}+\frac{ \theta^b_{{\v_1}}-\theta^b_{{\v_2}}}{2\pi n}\right)
\end{align}
where ${\v_1}$ and ${\v_2}$ are the two endpoints of link ${\l_0}$, $\mathcal{N}' = \frac{\sqrt{\pi}}{nW_a}\cdot (2\pi n)^2$ is a normalization factor, and we have neglected the infinitesimal background field. Integrating over $\theta^b_{{\v_2}}$ yields
\begin{align}
   \mathcal{B} &=  2\sum_{z^b_{{\l_0}}}  \int_0^{2\pi n} d\theta^b_{{\v_1}} \left(-pq+\frac{kq^2}{2n}\right)z^b_{\l_0} \left(\mathrm{e}^{-\left(z^b_{\l_0}-1+\frac{\theta^b_{{\v_1}}}{2\pi n}\right)^2 n^2W_a^2}-\mathrm{e}^{-\left(z^b_{\l_0}+\frac{\theta^b_{{\v_1}}}{2\pi n}\right)^2 n^2W_a^2}\right) / \mathcal{N}' \nonumber\\[.2cm]
    &=  (4\pi n) \left(-pq+\frac{kq^2}{2n}\right) \int_{-\infty}^{\infty} dz \ \mathrm{e}^{-z^2 n^2W_a^2} / \mathcal{N}' \ = \ \left(-\frac{pq}{ \pi n}+\frac{kq^2}{2\pi n^2}\right)
\end{align}
where in the second line we have used combined the summation over $z^b_{{\l_0}}$ and integral over $\theta^b_{{\v_1}}$ to make up an infinite integral for $\bar{b}$. We thus find a constant Berry curvature all over the domain of integral. The Chern number is thus $2\pi n^2 \mathcal{B} =  -2pqn+kq^2$.

\bibliography{ref}
\bibliographystyle{JHEP}

\end{document}